\documentclass{jpp}
\usepackage{graphicx}
\usepackage{hyperref}
\usepackage{caption}
\usepackage{subcaption}
\usepackage{xcolor}
\usepackage{multirow}
\usepackage{bm}

\usepackage[utf8]{inputenc}
\usepackage[T1]{fontenc}
\usepackage{amsmath}

\shorttitle{Extending near-axis equilibria in DESC}
\shortauthor{D. Panici, E. Rodriguez}

\author{Dario Panici\aff{1}
  \corresp{\email{dpanici@princeton.edu}},
  Eduardo Rodriguez\aff{2}, Rory Conlin\aff{3}, Daniel Dudt\aff{4}, Egemen Kolemen \aff{1}\corresp{\email{ekolemen@princeton.edu}}}

\affiliation{\aff{1}Princeton University, Princeton, New Jersey 08544
\aff{2}Max Planck Institute for Plasma Physics, 17491 Greifswald, Germany
\aff{3}Institute for Research in Electronics and Applied Physics, University of Maryland, College Park, MD 20742, USA
\aff{4} Thea Energy, USA
}

\title{Extending near-axis equilibria in \texttt{DESC}}
\date{April 2023}

\begin{document}

\maketitle

\begin{abstract}
    The near-axis description of optimised stellarator fields has proven to be a powerful tool both for design and understanding of this magnetic confinement concept. The description consists of an asymptotic model of the equilibrium in the distance from its centermost axis, and is thus only approximate. Any practical application therefore requires the eventual construction of a global equilibrium. This paper presents a novel way of constructing global equilibria using the \texttt{DESC} code that guarantees the correct asymptotic behaviour imposed by a given near-axis construction. The theoretical underpinnings of this construction are carefully presented, and benchmarking examples provided. This opens the door to an efficient coupling of the near-axis framework and that of global equilibria for future optimisation efforts.
\end{abstract}

\section{Introduction}
Stellarators offer a promising path to realizing controlled thermonuclear fusion as steady-state magnetic plasma confinement schemes \citep{boozer2021stellarators}. Their benefits stride from the freedom in the three-dimensional shaping of the magnetic field, away from the constraining requirement of \textit{axisymmetry} \citep{wesson2011tokamaks}, which is at the heart of the \textit{tokamak} concept. However, this increase in the complexity of the magnetic field shaping and the loss of symmetry requires careful tailoring, in particular, to prevent prompt collisionless losses of particles. The latter is commonly achieved through numerical optimisation, by seeking properties such as \textit{quasisymmetry} \citep{boozer_transport_1983,nuhrenberg_quasi-helically_1988, rodriguez2020necessary} or, more generally, \textit{omnigeneity} \citep{bernardin1986,Cary1997,hall1975,landreman_omnigenity_2012, dudt_magnetic_2024}, alongside other additional requirements (aspect ratio, magnetohydrodynamic (MHD) stability, coil complexity, \textit{etc}). The problem of finding stellarator designs is therefore a multi-objective optimization problem in a high-dimensional optimization space, with multiple minima and in which evaluation of each point (i.e. computation of 3D isotropic pressure, ideal magnetohydrostatic (MHS) equilibrium) is time consuming. While this method has been successfully applied to produce attractive stellarator designs \citep{landreman_magnetic_2022, nuhrenberg_quasi-helically_1988, anderson_helically_1995, ku_new_2010, zarnstorff_physics_2001}, it is a daunting (if not impossible) task to conduct an exhaustive solution space scan \citep{landreman_optimized_2019, giuliani2024comprehensive}.
\par
To alleviate some of this burden, and gain additional insight on what that vast space of stellarators may look like, analytic insight can be gained through the near-axis expansion (NAE) \citep{mercier_equilibrium_1964,Solovev1970,lortz1976equilibrium,garren_magnetic_1991,landreman_constructing_2019,rodriguez2024near}. This is an asymptotic description of the field in the distance from its centre (the magnetic axis), which constitutes a simple consistent model of stellarator fields. This approach has the benefit of reducing the MHS equilibrium problem, alongside omnigeneity conditions, to a reduced set of parameter and functions \citep{garren_existence_1991,landreman_direct_2019,plunk_direct_2019,rodriguez2024near} related through algebraic and ordinary differential equations that may be numerically solved orders of magnitude faster than their global counterparts \citep{landreman_direct_2019,panici_desc_2023}. This has enabled more exhaustive exploration of the space of stellarators \citep{landreman_mapping_2022, rodriguez_constructing_2023, giuliani2024comprehensive}. Besides this practical difference, NAE theory has also been shown to be a natural lens through which to understand the space of optimised stellarators, particularly quasisymmetric ones \citep{landreman_constructing_2019,rodriguez_phases_2022,rodriguez_constructing_2023}. 
\par
The NAE theory offers a sensible path to searching for candidate stellarator configurations with desirable confinement properties. However, the theory is still fundamentally an asymptotic theory: the solutions it finds are only valid in some volume near the axis. The further away, the less reliable the description becomes, and thus global MHS solutions are still required to fully characterize the stellarator equilibrium. It is natural, though, to employ the near-axis construction as a starting point for higher-fidelity field construction, and thus it is necessary to devise a way of connecting global equilibria solutions to the near-axis ones. The current standard method evaluates the NAE field at a finite distance from the axis, constructs the corresponding flux surface and uses it as an input to conventional nested flux surface global (fixed-boundary) MHS codes \citep{DESC,hirshman_steepestdescent_1983,hindenlang_computing_2025}. Although this procedure has proved effective, there is a blatant fault to it: the NAE is being used to inform the global solution at a point where it is least valid (far from the axis). The result is that many of the features curated into the NAE fields are lost when building the equilibria, especially at lower aspect ratios. 
\par
This paper introduces a new method of finding global 3D MHS equilibrium solutions, in the equilibrium solver \texttt{DESC}, which are intimately connected to NAE theory. We do so by constraining the global solution's near-axis behavior directly, using information from the NAE where it is most valid. This way of connecting the asymptotic behaviour to global equilibrium shall preserve any near-axis property studied, understood or optimised for within the near-axis framework. While we do so, we leave freedom to the global MHS solver to mold the solution far from axis, where the NAE loses accuracy. This results in global MHS solutions with near-axis behavior consistent with the NAE, and far-from-axis behavior consistent with solving the global MHS equilibrium equations. 
\par
The paper is organised as follows. Section~\ref{sec:nae-constraints-derivation} presents both formulations of the equilibrium problem as they are done in the NAE and \texttt{DESC}, in order to link them to each other. Section~\ref{sec:flux_surfs} will then use this connection to derive, order-by-order up to second order in the distance from the axis, the geometric constraints on the \texttt{DESC} equilibrium which enforce the NAE behavior. Section~\ref{sec:NAE_verification} then presents some examples of the constrained equilibria, which are also used to verify the correct enforcement of the NAE behavior in the global equilibrium.

\section{Connecting the near-axis description}\label{sec:nae-constraints-derivation}
The goal of the paper is simple: we aim to construct global equilibria given some prescribed near-axis behaviour. To do so, we first need to understand how to deal with the equilibrium problem. 

\subsection{The equilibrium problem}
We define an equilibrium solution as a magnetic field with nested flux surfaces that satisfies the MHS equilibrium equation $\mathbf{j}\times\mathbf{B}=\nabla p$, where $\mathbf{j}=\mu_0^{-1}\nabla\times\mathbf{B}$ is the plasma current density and $p$ is the plasma pressure. The magnetic field must, of course, be a solenoidal vector field, $\nabla\cdot\mathbf{B}=0$, and we define a toroidal magnetic flux $2\pi\psi$ whose level sets define nested flux surfaces satisfying $\mathbf{B}\cdot\nabla\psi=0$. Both the global equilibrium solver and the near-axis description attempt the construction of equilibrium fields of this form. 
\par
Within this set of equations, we confer the latter two a more prominent role. That is, we prioritize the magnetic field being solenoidal and having nested flux surfaces, which we impose on the field in an \textit{exact} form. This is done by adopting the \textit{inverse-coordinate} form of the problem \citep{hirshman_steepestdescent_1983, bauer_computational_1978}. Instead of describing the magnetic field $\mathbf{B}$ explicitly as a function of space, we instead focus on describing the magnetic flux surfaces on which the field lives, as well as the form in which field lines wrap over them. 
\par
To construct such a description, we start by writing the magnetic field in its Clebsch form \citep{d2012flux},
\begin{equation}
    \mathbf{B}=\nabla\psi\times(\nabla\vartheta-\iota\nabla\varphi), 
\end{equation}
where $\vartheta$ and $\varphi$ are poloidal and toroidal angles (in straight field line coordinates with non-vanishing Jacobian), and $\iota$ (a function of $\psi$) is the rotational transform. A field written in this form satisfies both field requirements exactly. Defining a position vector $\mathbf{x}=\mathbf{x}(\psi,\vartheta,\varphi)$, and taking that set of straight field line coordinates as our set of independent coordinates, the magnetic field may then be expressed as,
\begin{equation}
    \mathbf{B}=\mathcal{J}_\vartheta^{-1}\left(\frac{\partial \mathbf{x}}{\partial\varphi}+\iota\frac{\partial\mathbf{x}}{\partial\vartheta}\right), \label{eqn:Binx}
\end{equation}
where $\mathcal{J}_\vartheta=\partial_\psi\mathbf{x}\times\partial_\vartheta\mathbf{x}\cdot\partial_\varphi\mathbf{x}$ is the Jacobian of the straight field line coordinate system. Full-knowledge of flux surfaces in straight field line coordinates (i.e. $\mathbf{x}$), along with the rotational transform, uniquely defines the magnetic field. Alternatively, one could provide information about the average toroidal current rather than the rotational transform (see Appendix~\ref{app:iota-current-formula}). 
\par

The central task is then to find a function $\mathbf{x}(\psi,\vartheta,\varphi)$, which we may represent in a multitude of ways. The near-axis expansion (NAE), in its inverse-coordinate form pioneered by \cite{garrenboozer1991a}, takes as a reference of this $\mathbf{x}$ the magnetic axis. This is a natural choice given that the NAE is concerned with providing a consistent asymptotic description of the field near the magnetic axis (i.e., $\psi=0$). Explicitly, 
\begin{equation}
\mathbf{x}(\psi,\vartheta,\varphi)=\mathbf{r}_0+X(\psi,\vartheta,\varphi)\hat{\pmb\kappa}+Y(\psi,\vartheta,\varphi)\hat{\pmb\tau}+Z(\psi,\vartheta,\varphi)\hat{\pmb t}, \label{eqn:xnae}
\end{equation}
where $\mathbf{r}_0$ describes the magnetic axis, and the vector triad $\{\hat{\pmb\kappa},\hat{\pmb\tau},\hat{\pmb t}\}$ corresponds to the Frenet-Serret basis \citep{frenet1852courbes, animov2001differential} (namely, the normal, binormal and tangent vectors defined with respect to the magnetic axis). Here we specialise to regular magnetic axes with no, or only few, isolated flattening points (i.e. vanishing curvature points) in which such a frame (or its signed version \citep{carroll2013improving,plunk_direct_2019, mata2023helicity, rodriguez2024near}) exists. This is enough to describe the asymptotic behaviour of all optimised configurations; namely, quasisymmetric and quasi-isodynamic ones \citep{gori1997quasi,plunk_direct_2019}. In the near-axis expansion, then, finding a magnetic field corresponds to finding an asymptotic form of the functions $X,~Y$ and $Z$, which describe the shape of flux surfaces, alongside the rotational transform and the axis shape. This description is performed in a special class of straight field line coordinates, Boozer coordinates $\{\psi,\vartheta_\mathrm{B},\varphi_\mathrm{B}\}$, which eases the treatment of optimised stellarators due to the simple involvement of $|\mathbf{B}|$ \citep{boozer_plasma_1981}, which imposes additional constraints in the construction. 
\par
In the case of the global equilibrium description, the numerical solver \texttt{DESC}, which we shall focus on, describes flux surfaces not with respect to the axis (like new developments on other alternatives can \citep{hindenlang_computing_2025}), but in cylindrical coordinates so that,
\begin{equation}
    \mathbf{x}_\mathrm{DESC}=R \hat{\bm{R}} + Z \hat{\bm{z}}. \label{eqn:xDesc}
\end{equation}
It is then the task of the solver to find the appropriate $R$ and $Z$ functions as a function of $\{\psi,~\vartheta,~\varphi\}$. Given that the description is in cylindrical coordinates, the toroidal angle is naturally chosen to be the cylindrical one, $\phi$.\footnote{This choice, although natural given the representation, is actually not the most convenient in many instances in which the configurations are significantly shaped and twisted.} Because we insist on this specific form of the toroidal angle, it requires a very particular choice of the poloidal angle $\vartheta$ in order to guarantee the coordinate system to be a straight field line coordinate system. This coordinate system is known as PEST coordinates $\{\psi,\vartheta_\mathrm{PEST},\phi\}$. However, the global solver does not know at the starting point what $\vartheta_\mathrm{PEST}$ is, as this depends on the final magnetic field solution. Hence, in practice, the solver must introduce a stream function $\lambda$ such that $\vartheta_\mathrm{PEST}=\theta+\lambda$, where $\theta$ can be any arbitrary poloidal angle used by the equilibrium solver. In that case, the magnetic field, as in Eq.~(\ref{eqn:Binx}) is written as,
\begin{subequations}
    \begin{equation}
\mathbf{B}=\mathcal{J}_\theta^{-1}\left[(1+\partial_\theta\lambda)\frac{\partial \mathbf{x}}{\partial\phi}+(\iota-\partial_\phi\lambda)\frac{\partial \mathbf{x}}{\partial\theta}\right]. \label{eqn:Binx_lambda}
    \end{equation}
\end{subequations}
In the case of the global solver then, a magnetic field is uniquely described by finding functions $R,~Z$ and $\lambda$ as a function of $\{\psi,~\theta,~\phi\}$. 
\par
So far, we have only represented a magnetic field but have not yet solved the equilibrium. We must still find a consistent set of flux surfaces $\mathbf{x}$ (described in appropriate coordinates), such that the resulting magnetic field is in equilibrium. We must deform flux surfaces until the field comes to equilibrium; \textit{i.e.}, it minimises the force residual $\mathbf{F}=\mathbf{j}\times\mathbf{B}-\nabla p$ for a prescribed function $p(\psi)$. In the context of the near-axis expansion, the consistent construction of flux surfaces and Boozer coordinates can be carried out systematically by setting $\mathbf{F=0}$ order-by-order, and detailed descriptions of how to do this may be found in \cite{landreman_constructing_2019} and \cite{rodriguez2024near}. We content ourselves with knowing that, at the end of the day, we will be able to have an asymptotic form of $X,Y,Z$, rotational transform and the axis shape. 
\par
The equilibrium solver \texttt{DESC} seeks equilibria by numerically and iteratively minimising the force residual $\mathbf{F}$ directly. Note that this is not the only existing practical approach, with the important case of the \texttt{VMEC} code approaching the problem by minimising energy \citep{hirshman_steepestdescent_1983}. To be more explicit about the force residual, consider first the covariant form of the magnetic field,
\begin{equation}
\mathbf{B}=B_\theta\nabla\theta+B_\phi\nabla\phi+B_\psi\nabla\psi. \label{eqn:covB}
\end{equation}
There might appear to be new information in the problem, but all of the covariant components in Eq.~(\ref{eqn:covB}) can be directly computed from the geometry by simply leveraging the \textit{dual relations} \citep[Eqs.~(2.3.12)-(2.3.13)]{d2012flux} $\nabla q_i=\epsilon_{ijk}\partial_{q_j}\mathbf{x}\times\partial_{q_k}\mathbf{x}$ and Eq.~(\ref{eqn:Binx_lambda}). With this, the force residual is (see \cite{panici_desc_2023} as well),

\begin{multline}
    \mathbf{F}=\left[\mathcal{J}_\theta^{-1}\mu_0^{-1}(\partial_\phi B_\psi-\partial_\psi B_\phi)(1+\partial_\theta\lambda)-\mathcal{J}_\theta^{-1}\mu_0^{-1}(\iota-\partial_\phi\lambda)(\partial_\psi B_\theta-\partial_\theta B_\psi)-p'\right]\nabla\psi\\
    -\mathcal{J}_\theta^{-1}\mu_0^{-1}(\partial_\theta B_\phi-\partial_\phi B_\theta)((1+\partial_\theta\lambda)\nabla\theta-(\iota-\partial_\phi\lambda)\nabla \phi).
    \label{eqn:DESCforceBalance}
\end{multline}

The $\nabla\psi$ component can be interpreted as the radial force balance equation, with the helical component representing $\mathbf{j}\cdot\nabla\psi$. Thus, the global solver is set to find $\{R,~Z,~\lambda\}$ as a function of $\{\psi,\theta,\phi\}$ that make Eq.~(\ref{eqn:DESCforceBalance}) vanish. 
\par
The near-axis and global treatment of the problem thus have a key difference not only in the coordinate representation used, but also in their treatment of boundary conditions. One must provide certain inputs to the equilibrium problem to describe different equilibria. In the NAE case, this prescription occurs outwards; that is, one must provide the shape of the axis, and then, order-by-order, some features of the flux surfaces away from it. Which features must be provided depends on the type of stellarator under consideration, and the order of the expansion of interest (see \cite{landreman_constructing_2019} and \cite{rodriguez2024near}). The global solution fundamentally attempts the opposite: one specifies the shape of the boundary, and the solution is constructed inwards (what is known as a \textit{fixed-boundary} solution). This is most straightforwardly understood by imagining a vacuum field in the form of Laplace's equation. From this difference it should be clear that there is not a one-to-one correspondence between the two approaches. A finite truncated near-axis construction does not in principle uniquely describe a global field solution, but rather simply constrains its core behaviour. We are now in a position to connect both. 

\subsection{Near-axis expansion in \texttt{DESC}}
In the near-axis expansion we learned that the description of flux surfaces $\mathbf{x}(\psi,\vartheta_B,\varphi_B)$ is given by $\{X,Y,Z\}$, Eq.~(\ref{eqn:xnae}). Their form is particularly simple, as may be directly appreciated by considering their Taylor-Fourier expansion \citep{garrenboozer1991a,landreman_constructing_2019},
\begin{equation}
    f(\psi,\theta_\mathrm{B},\varphi_\mathrm{B})=\sum_{l=0}^\infty r^l\sum_{m=0}^l{}^{'}\left(f_{lm}^C(\varphi_\mathrm{B})\cos m\vartheta_\mathrm{B}+f_{lm}^S(\varphi_B)\sin m\vartheta_\mathrm{B}\right), \label{eqn:fNAE}
\end{equation}
where $r=\sqrt{2\psi/\bar{B}}$ is a pseudo-radial coordinate, $\bar{B}$ is a representative normalisation value of the magnetic field, and the sum $\sum'$ denotes sum over even or odd orders depending on the parity of $l$. Because the poloidal dependence grows in a controlled manner linked to the powers of $r$, the flux surfaces gain complexity order-by-order; they start by being elliptical, then acquire triangularity, and so on.  
\par
To impose this asymptotic form of the field on the global equilibrium solver, we must be able to relate this asymptotic description to the representation in \texttt{DESC}, and constrain the latter according to the former. The structure in Eq.~(\ref{eqn:fNAE}) is a consequence of maintaining the field description singularity-free close to the magnetic axis ($\psi=0$) \citep{kuo1987numerical,garren_magnetic_1991,landreman_direct_2018}. This requirement was indeed observed in the implementation of \texttt{DESC}, where the global solver employs a Zernike basis to represent $\{R,Z,\lambda\}$, crucially coupling the powers of \texttt{DESC}'s chosen pseudo-radial coordinate $\rho=\sqrt{\psi/\psi_\mathrm{edge}}$ to the poloidal modes $\theta$. This makes the treatment in \texttt{DESC} close to the form used in the near-axis description. 
\par
To make that connection, though, we must consider the differences carefully. One is the use of different re-scaled pseudo-radial coordinates; we may relate $\rho$ to the NAE $r$ by $\rho=r\sqrt{\bar{B}/2\psi_\mathrm{edge}}$. The second point of difference is the use of a different poloidal angle. While the near-axis description uses the poloidal Boozer angle, $\vartheta_\mathrm{B}$, the global equilibrium is described in a general poloidal angle $\theta$. The simplest comparison between the two descriptions is then made when $\theta$ is forced to match the Boozer angle, at least to leading order near the axis. When imposing the near-axis constraints on $\theta$ as if it were Boozer coordinates, the solver is forced to use a very particular form of $\lambda$, and thus can be overly constraining. Defining $\nu=\varphi_\mathrm{B}-\phi$ as the function that maps the Boozer toroidal angle to the cylindrical one, our choice of $\theta$ dictates that
    \begin{gather}
        \lambda=-\iota\nu. \label{eqn:lam_nu_Booz}
    \end{gather}
This simplest choice of poloidal angle could be modified to accommodate a more general and efficient $\theta$. Details about how this would be incorporated are shown in Appendix~\ref{app:desc-theta-constraint}, but we shall nevertheless consider the simplest scenario in the main body of the paper, which should suffice as a proof of principle.
\par
With the coordinates in place, what remains from the comparison is the different radial-poloidal structure of the NAE expansion and the Zernike basis. In the equilibrium solver a generic function is written as (ignoring the toroidal coordinate for now),
\begin{equation}
    f(\rho,\theta)=\sum_{l=0}^\infty\sum_{m=-l}^l{}^{'}f_{lm}\mathcal{Z}_l^m(\rho,\theta), \label{eqn:fZer}
\end{equation}
where the $\sum^{'}$ again denotes a sum over the same parity as $l$, and the Zernike polynomials are,
\begin{equation}
    \mathcal{Z}_l^m(\rho,\theta)=\mathcal{R}_l^{|m|}\begin{cases}
        \cos(m\theta) & m\geq0, \\
        \sin(|m|\theta) & m<0,
    \end{cases}
    \label{eqn:zern}
\end{equation}
where the radial parts are the shifted Jacobi polynomials, given as,
\begin{equation}
    \mathcal{R}_l^{|m|}=\sum_{k=0}^{(l-|m|)/2}\frac{(-1)^k(l-k)!}{k!\left(\frac{l+m}{2}-k\right)!\left(\frac{l-m}{2}-k\right)!}\rho^{l-2k}. \label{eqn:Rlm}
\end{equation}
The coefficients $\{f_{lm}\}$ for $\{R,Z,\lambda\}$ are the set which \texttt{DESC} solves for in order to construct the equilibrium field. To make a one-to-one connection to the NAE, Eq.~(\ref{eqn:fNAE}), we need to unravel this sum and pick the different powers of $\rho$. Generally, a function $f$ defined by $f_{lm}$ coefficients can be written in terms of the equivalent near-axis coefficients as
\begin{equation}
    f_{l|m|}^{C/S}=\left(\frac{\rho}{r}\right)^l\times\begin{cases}
    \begin{aligned}
        &(-1)^{l/2}\sum_{k=l/2}^\infty (-1)^k\binom{k+l/2}{l}\binom{l}{(l-|m|)/2}f_{2k,m} \quad |m| \mathrm{~is~even} \\
        &(-1)^{(l+1)/2}\sum_{k=(l+1)/2}^\infty (-1)^k\binom{k+(l-1)/2}{l}\binom{l}{(l-|m|)/2}f_{2k-1,m} \quad |m| \mathrm{~is~odd}, \\
        \end{aligned}
    \end{cases} \label{eqn:generalRelZerNAE}
\end{equation}
and the sign of $m$ has been taken to represent, as is the case in the \texttt{DESC} notation, the sine or cosine terms $S/C$ (negative and positive respectively).\footnote{Of course, in practice the sum is not over an infinity of terms, but rather a finite number of them, depending on the resolution used.} The above only holds for $l\geq |m|$, and $m$ sharing the same parity as $l$. Otherwise, the contribution will be zero. Note how for large $k$ at constant $l$ and $m$, the factor in front of the Zernike components goes like $\sim k^l$. This shows the exponential contribution from the higher-order Zernike modes, which must therefore be conveniently truncated. See details on the derivation in Appendix~\ref{app:zern_to_nae}. Thus, imposing the near-axis behaviour constitutes imposing constraints on linear combinations of the \texttt{DESC} degrees of freedom.
\par
As a final remark, we note that what have been referred to as coefficients $f_{lm}$ so far are generally functions of $\phi$. In the pseudo-spectral representation of \texttt{DESC}, we may then treat the toroidal dependence with a Fourier series in $\phi$, with the $n$ Fourier components corresponding to $f_{lmn}$, with $n\geq0$ representing cosines, and $n<0$ sines. 

\section{Constraining the global equilibrium}\label{sec:flux_surfs}
Equations (\ref{eqn:generalRelZerNAE}) define a linear mapping between the Zernike and the NAE basis. If the latter is specified, then, the above constitutes a constraint on a linear combination of Zernike components. It might thus seem that there is nothing else that needs to be done in order to bridge the NAE and the global equilibrium solver. However, it is not the full story.
\par
As discussed earlier, the description of flux surfaces, the central element in the description of the field, assumed a different form in the standard near-axis approach to equilibrium versus how the solver \texttt{DESC} treats the problem, Eqs.~(\ref{eqn:xnae}) and (\ref{eqn:xDesc}). The latter employs a cylindrical geometry in real space $R(\rho,\theta,\phi)$ and $Z(\rho,\theta,\phi)$, where $\phi$ \textbf{must} be the geometric cylindrical angle. This we may refer to as a description in the `lab-frame' (independent of the field). The description in the near-axis expansion, instead, is intimately linked to the form of the field itself, as surfaces are described with respect to the magnetic axis in its Frenet-Serret frame, and in Boozer coordinates. This implies that shapes described in the near-axis expansion need to be appropriately transformed into the `lab-frame' in order to impose the constraint on the field appropriately.
\par

\subsection{Zeroth order: the magnetic axis}
To leading order in the near-axis description, flux surfaces reduce to a single closed curve at the origin ($\psi=0$): the magnetic axis. In practice, such a closed curve is described as a set of Fourier harmonics (in $\phi$) in cylindrical coordinates $\{R,~Z\}$. That is,
\begin{subequations}
    \begin{gather}
        R=R_0+\sum_{n=1}^N (R_{n}^C\cos n\phi+R_{n}^S\sin n\phi), \\
        Z=\sum_{n=1}^N (Z_{n}^C\cos n\phi+Z_n^S\sin n\phi),
    \end{gather}
\end{subequations}
where the Fourier coefficients are constant. This form of describing the axis is natural to NAE codes (such as \texttt{pyQSC}), and equilibrium (such as \texttt{DESC}). Following the discussion on the Zernike basis, using Eq.~(\ref{eqn:f00}), the relevant constraint for the Zernike modes are then,
\begin{subequations}
\begin{gather}
    R_{n}^{C/S}=\sum_{k=0}^\infty (-1)^k R_{2k,0,\pm|n|}, \\
    Z_{n}^{C/S}=\sum_{k=0}^\infty (-1)^k Z_{2k,0,\pm|n|}.
\end{gather}\label{eqn:rAxisZern}
\end{subequations}
We shall thus have $2(2N+1)$ constraints to impose onto the Zernike harmonics in order to constrain the magnetic axis to match that of the NAE. No further geometric consideration is needed here. 

\subsection{First order: elliptical surfaces}\label{sec:first-order-geom}
At first-order things start to become more interesting, and we need to explicitly transform the constraints into cylindrical coordinates (the language of \texttt{DESC}) using the information in our magnetic-axis frame. This includes transforming the Boozer toroidal angle to the cylindrical angle $\phi$. This transformation is routinely performed numerically by codes like \texttt{pyQSC} to represent near-axis magnetic fields in the lab-frame, and was previously explored in \cite{landreman_direct_2018}. Here we present an alternative simple geometric treatment of the problem. 
\par
\begin{figure}
    \centering
    \includegraphics[width=0.4\textwidth]{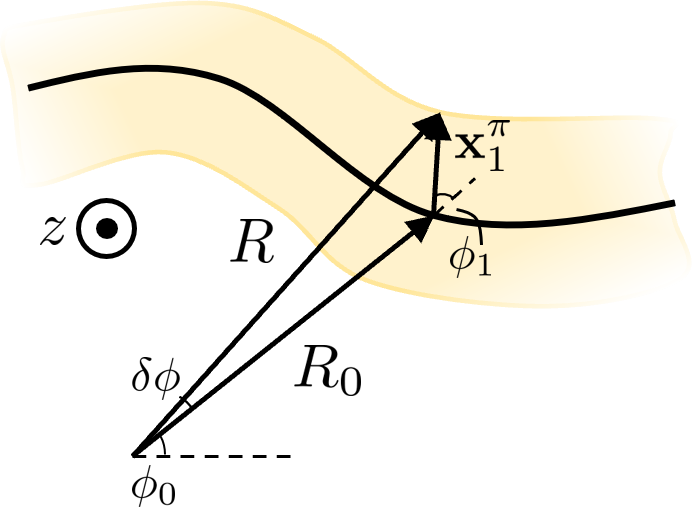}
    \caption{\textbf{Diagram illustrating the key element for the geometric transformation at first order.} Schematic diagram showing the position of a point in the surface (at radial distance $R$ and angle $\phi_0+\delta\phi$), in reference to other quantities. These include the position along the magnetic axis (radial position $R_0$ and angle $\phi_0$), and the $\rho\mathbf{x}_1$ from the axis to the point, projected onto the $R,~\phi_c$ plane (superindex $\pi$).  }
    \label{fig:diagramGeoZernNAE}
\end{figure}

Start by defining the flux surface as $\mathbf{x}=\mathbf{r}_0+\mathbf{x}_1$, where $\mathbf{x}_1\propto \rho$ and thus is in the asymptotic sense small. That is, we construct flux surfaces near the axis through a small displacement $\mathbf{x}_1$ from it. Our task is to find $R$ and $Z$ corresponding to these points. Start with $R$, the major radius, and consider the projection of the problem to a constant $Z$ plane, which we define as the $\pi$ plane (see the diagram in Figure~\ref{fig:diagramGeoZernNAE}). 
Define a point on the magnetic axis by $R_0$ (the radial distance to a point along the axis), and $\phi_0$ (the cylindrical angle of said point). The displacement $\mathbf{x}_1$ is then a function of that point (see Fig.~\ref{fig:diagramGeoZernNAE}), whose projection normal to $z$ we define to be $\mathbf{x}_1^\pi$. Define $\phi_1$ to be the angle that $\mathbf{x}_1^\pi$ makes with $\hat{\bm{R}}_0$, which satisfies $\cos\phi_1=\hat{\bm{R}}_0\cdot\mathbf{x}_1^\pi/|\mathbf{x}_1^\pi|$. Here $\hat{\bm{R}}_0$ corresponds to the radial vector direction evaluated at the axis; that is, $\hat{\bm{R}}_0=\hat{\bm{R}}(\phi=\phi_0$). We then define $\delta \phi$ to be the angle which, when added to $\phi_0$, results in the cylindrical toroidal angle of the point on the flux surface $\mathbf{x}_0+\mathbf{x}_1$.
\par 
Application of the sine rule gives,
\begin{equation}
    \delta\phi\approx\frac{x_1^\phi}{R_0}, \label{eqn:deltaphi1}
\end{equation}
where $x_1^\phi=|\mathbf{x}_1^\pi|\sin\phi_1$ and only the leading piece in $\rho$ is being kept (so, $R \approx R_0$) and $\delta \phi \ll 1$ has been assumed. We define $x_1^j=\mathbf{x}_1\cdot\hat{j}$ for $j=R,~\phi,~z$ as the projections of $\mathbf{x}_1$ along the cylindrical basis vectors, which when expressed in terms of the Frenet-Serret frame of the axis will involve various projections of the triad $\{\hat{\pmb\kappa},\hat{\pmb\tau},\hat{\pmb t}\}$. 
\par
Applying the cosine rule, and defining $R-R_0=\delta R$, we find that to leading order in $\rho$,
\begin{equation}
    \delta R\approx x_1^R.
\end{equation}
 Therefore, the radial position as a function of the geometric angle at the point on the surface,\footnote{All the quantities we found before can be regarded as functions of the point along the magnetic axis. Thus, the use of $\phi-\delta\phi$ for the appropriate description.} is given as
\begin{equation}\label{R_exp_1st_oder}
    R(\phi)\approx R_0(\phi-\delta\phi)+\delta R(\phi-\delta\phi)\approx R_0(\phi)+\left[\delta R(\phi)-\delta\phi\partial_\phi R_0\right]+O(\rho^2),
\end{equation}
where all the quantities in brackets are evaluated at $\phi$, and $R_0$ is a known function (from the shape of the magnetic axis). The quantity in square brackets is what will become the constraint on the Zernike basis, and in the asymptotic sense is correct to $O(\rho)$. 
\par
 Then, plugging in our expressions for $\delta \phi$ and $\delta R$, the first order $\rho$ terms in $R$ are,
\begin{equation}
    R_1=x_1^R-x_1^\phi\frac{\partial_\phi R_0}{R_0}. \label{eqn:R1}
\end{equation}
This form preserves the simple $\theta$-harmonic content of $\mathbf{x}_1$, which from the near axis has the simple form $\mathbf{x}_1=\cos\theta(X_1^c\hat{\pmb\kappa}+Y_1^c\hat{\pmb\tau})+\sin\theta (X_1^s \hat{\pmb\kappa} + Y_1^s\hat{\pmb\tau} )$. Thus, Eq.~(\ref{eqn:R1}) can be written as $R_1=\mathcal{R}_{1,1}(\phi)\cos\theta+\mathcal{R}_{1,-1}(\phi)\sin\theta$, where the form of the coefficients follows directly from Eq.~(\ref{eqn:R1}) and the form of $\mathbf{x}_1$ (these are derived explicitly in Appendix~\ref{app:1st-order-coeff-derivation}). The subscripts in $\mathcal{R}_{nm}$ refer to the radial and poloidal order of the term, respectively. To evaluate these we need the projections of the axis normal and binormal onto the cylindrical basis, information which is known in the context of the near-axis expansion (numerically calculated by codes like \texttt{pyQSC}).
\par
For $Z$, analogously, we get 
\begin{equation}
    Z_1 = x_1^z-x_1^\phi\frac{\partial_\phi Z_0}{R_0}, \label{eqn:Z1}
\end{equation}
where $Z_0(\phi)$ is the $Z$ of the magnetic axis that is known fully as a function of $\phi$. Once again, one may collect terms to construct $Z_1=\mathcal{Z}_{1,1}(\phi)\cos\theta+\mathcal{Z}_{1,-1}(\phi)\sin\theta$.
\par
Using Eq.~(\ref{eqn:f11}), we may (using \texttt{DESC} notation) write,
\begin{subequations}
    \begin{gather}
        \frac{r}{\rho}\mathcal{R}_{1,1,n} = -\sum_{k=1}^M(-1)^kkR_{2k-1,1,n}, \\
        \frac{r}{\rho}\mathcal{R}_{1,-1,n} = -\sum_{k=1}^M(-1)^kkR_{2k-1,-1,n},
    \end{gather} \label{eqn:firstOrdTransf}
\end{subequations}
for all $n\in[-N,N]$ representing Fourier components in $\phi$, and the negative sign corresponds to the sine $\phi$ components. The left hand sides represent the NAE components, which may be obtained through the near-axis expansion, and thus serve as constraint on the \texttt{DESC} Fourier-Zernike mode amplitudes, which appear in the right hand side. The constraints for $Z$ have exactly the same form but with $\mathcal{Z}$.

\par

\subsubsection{Straight-field line angle}\label{sec:SFL}
Recall that in order to complete the description of a field in \texttt{DESC}, it is important to find the stream function $\lambda$ that transforms the angle $\theta$ to the PEST poloidal coordinate. In the approach here, and given that the near-axis expressions in Boozer coordinates are being used as constraints, we have (at least asymptotically) $\theta=\vartheta_B$, and thus $\lambda=-\iota\nu$, Eq.~(\ref{eqn:lam_nu_Booz}). Hence, by constraining $R$ and $Z$ as above, and in order for \texttt{DESC} to find an equilibrium, $\lambda$ is in practice constrained.  
\par
To find what this $\lambda$ is expected to be, we need to find $\phi$, the cylindrical angle, as a function of $\phi_\mathrm{B}$ and $\vartheta_\mathrm{B}$ within the near-axis expansion. We may write $\phi=\phi^{(0)}(\varphi_B)+r\phi^{(1)}(\vartheta_B,\varphi_B)+\dots$, in which the leading order expression is \citep[Eq.~(A20)]{landreman_constructing_2019}, 
\begin{equation}
    \lambda^{(0)}=-\iota_0\int_0^\phi\left(\frac{B_0}{G_0}\frac{\mathrm{d}\ell}{\mathrm{d}\phi}-1\right)\mathrm{d}\phi', \label{eqn:lamba_0}
\end{equation}
where $\ell$ is the length along the magnetic axis, $B_0$ the magnetic field magnitude, $G_0$ the poloidal Boozer current and $\iota_0$ the rotational transform, all the zero subscripts indicating evaluation on axis. 
\par
The first correction $\phi^{(1)}$ is directly related to $\delta\phi$ in Eq.~(\ref{eqn:deltaphi1}), so that $\nu^{(1)}=-\delta\phi_1=-x_1^\phi/R_0$, parametrised with the cylindrical angle on axis. Parametrised in terms of the standard cylindrical angle it would read,
\begin{equation}
    \lambda^{(1)}=\iota_0\frac{x_1^\phi}{R_0}\left(1-\frac{\lambda^{(0)}{}'}{\iota_0}\right),
\end{equation}
where the prime denotes a toroidal derivative.

\subsubsection{Interpretation of the 1st order transformation}
The above might appear to be a purely formal artifact resulting from the change of frame, but it has a straightforward geometric interpretation. The expressions in Eqs.~(\ref{eqn:R1}) and (\ref{eqn:Z1}) show the difference in describing the shape of cross-sections in the near-axis frame (cutting flux surfaces normal to the magnetic axis) and the `lab-frame' (where the cuts are made at constant cylindrical toroidal angle). The elliptical cross-sections that live in the plane normal to the axis will generally have some inclination respect to the cylindrical basis, by virtue of the axis being non-planar. Hence, a cut in the lab-frame introduces an additional projection.
\par
\begin{figure}
    \centering
    \includegraphics[width=0.8\textwidth]{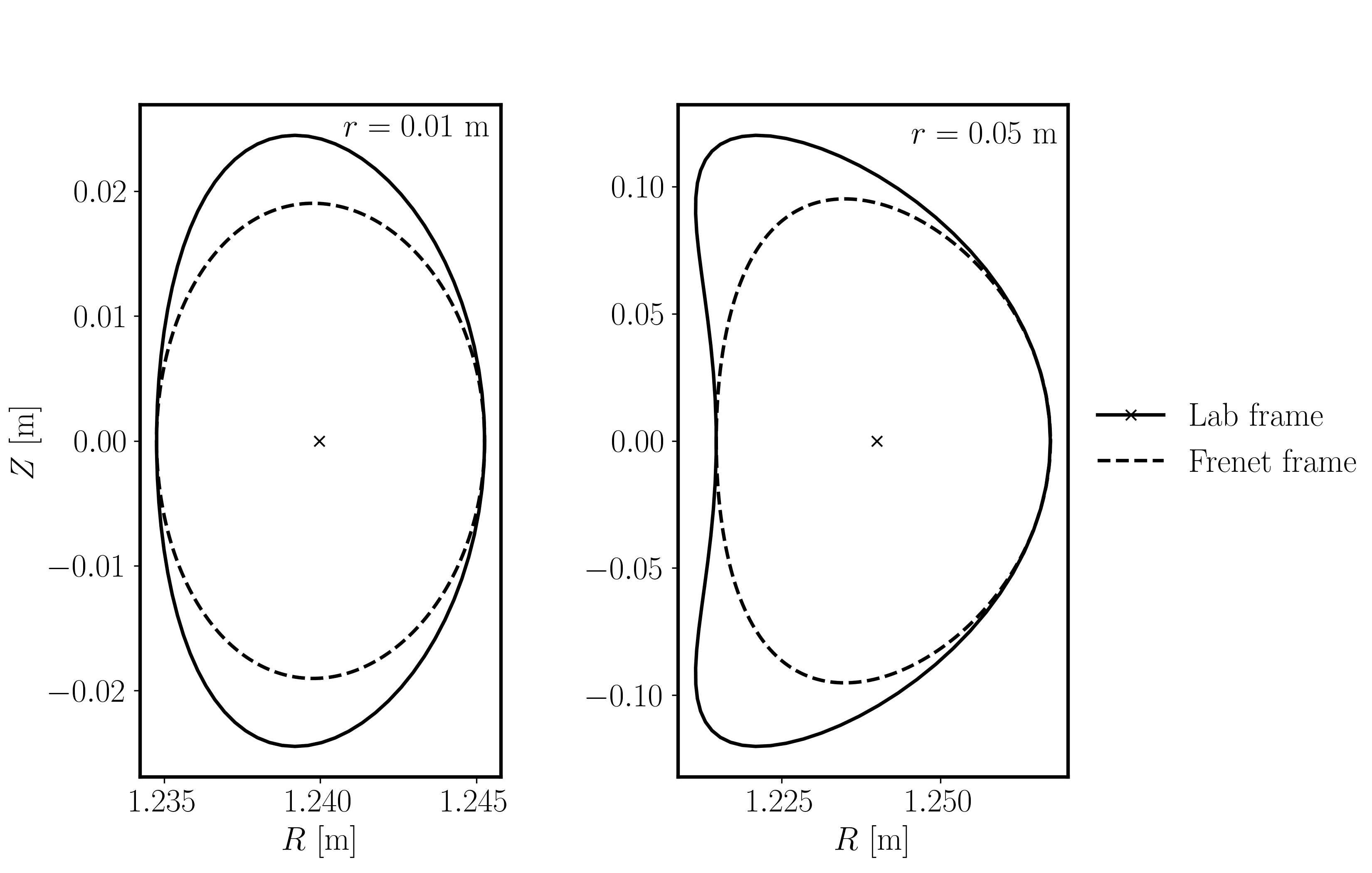}
    \caption{\textbf{Geometric deformation of cross-sections from the near-axis to the lab frame.} Example of the change in the cross-sections due to the geometric effects of going from the frame of the axis (broken lines) to the lab-frame (solid line). These correspond to the cross sections of the "precise QH" stellarator configuration from \cite{landreman_magnetic_2022} at one of its stellarator symmetric points. The left corresponds to the cross-section evaluated at $r=0.01$, while the shape to the right is at $r=0.05$. The shape to the left shows the enhancement of elongation in the vertical direction due to the inclination of the axis, and the right the change of triangularity but immutability of the centre of the cross-section. }
    \label{fig:geoDefCrossSec}
\end{figure}

Take as an illustrating example the cross-section at a stellarator symmetric point \citep{dewar_stellarator_1998}, which corresponds to an up-down symmetric ellipse in the plane normal to the magnetic axis. Because under the map $(\phi,\theta)\rightarrow(-\phi,-\theta)$ it must be the case that $R,Z\rightarrow R,-Z$, it follows that $R$ and $Z$ must be even and odd functions respectively. In particular, this means that $\partial_\phi R_0=0=\partial_\phi^2 Z_0$ at $\phi=0$. 
\begin{figure}
    \centering
    \includegraphics[width=0.3\textwidth]{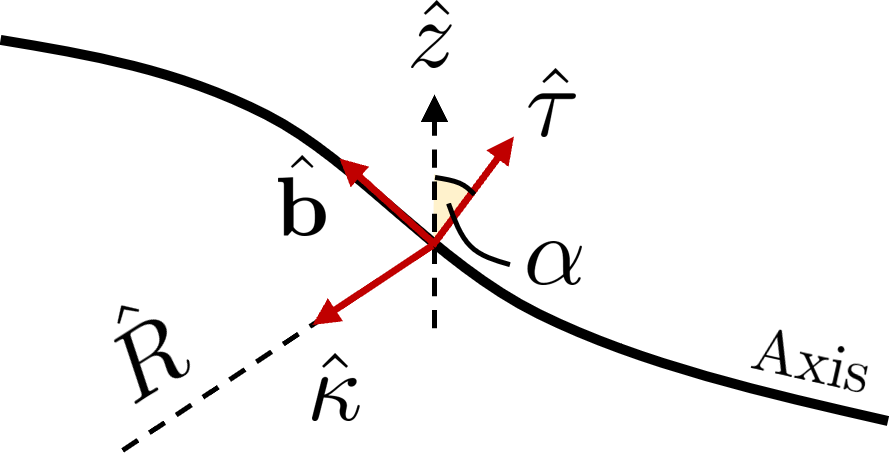}
    \caption{\textbf{Definition of the slant angle $\alpha$.} Diagram showing the definition of the angle $\alpha$ measuring the inclination of the magnetic axis at the origin ($\phi=0$) with the `lab' cylindrical coordinate system. The symbols have their usual meaning. }
    \label{fig:nuDef}
\end{figure}
Define a slant angle $\alpha$ as the angle that the plane normal to the magnetic axis makes with respect to the cylindrical toroidal angle; i.e., the magnetic axis is rotated about $\hat{R}$ by an angle $\alpha$ (see Figure~\ref{fig:nuDef}). Then we may write the components of $\hat{\pmb\tau}$ as $\tau^z=\cos\alpha$ and $\tau^\phi=\sin\alpha$, and because of the symmetry, $\hat{\kappa}=-\hat{R}$. The transformation in Eqs.~(\ref{eqn:R1}) and (\ref{eqn:Z1}) then reduces to a scaling of the ellipse in the binormal ($Y$) direction,
\begin{equation}
    \bar{Y}_1=Y_{1}\bigg(\tau^z-\underbrace{\frac{\partial_\phi Z_0}{R_0}}_{-\tau^\phi/\tau^z}\tau^\phi\bigg)=\frac{Y_{1}}{\cos\alpha}.
\end{equation}
Due to the inclination of the axis, the ellipticity of the cross-section changes (see an example in Figure~\ref{fig:geoDefCrossSec}), becoming stretched in the binormal direction. Away from this point of stellarator symmetry, the axis also presents an inclination respect to the radial coordinate, leading to an additional reshaping of the elliptical shape.
\par

\subsection{Second order: triangular surfaces}
The same approach detailed in the previous section for the first order can be extended to higher orders following the same geometric construction as that in Fig.~\ref{fig:diagramGeoZernNAE}. The details of how to proceed to second order are presented in Appendix~\ref{app:2nd-order-derivation}. Second order is perhaps the highest order of interest, given that near-axis constructions are hardly performed beyond it \citep{garren_existence_1991, landreman_constructing_2019,rodriguez2024near}.
\par
The derivation at second order proceeds by writing $\mathbf{x}=\mathbf{r}_0+\rho\mathbf{x}_1+\rho^2\mathbf{x}_2$, where $\mathbf{x}_2=X_2\hat{\pmb\kappa}+Y_2\hat{\pmb\tau}+Z_2\hat{\pmb t}$. Using the same considerations as in Fig.~\ref{fig:diagramGeoZernNAE}, 
\begin{subequations}
\begin{multline}
    R_2=-\frac{1}{2}\partial_\phi^2R_0\left(\frac{x_1^\phi}{R_0}\right)^2-\left(\frac{x_2^\phi}{R_0}-\frac{x_1^Rx_1^\phi}{R_0^2}\right)\partial_\phi R_0-\\
    -\frac{x_1^\phi}{R_0}\partial_\phi\left(x_1^R-\frac{x_1^\phi}{R_0}\partial_\phi R_0\right)
    +\left(x_2^R+\frac{(x_1^\phi)^2}{2R_0}\right). \label{eqn:R2}
\end{multline}
\begin{equation}
    Z_2=-\frac{1}{2}\partial_\phi^2 Z_0\left(\frac{x_1^\phi}{R_0}\right)^2-\left(\frac{x_2^\phi}{R_0}-\frac{x_1^Rx_1^\phi}{R_0^2}\right)\partial_\phi Z_0-\frac{x_1^\phi}{R_0}\partial_\phi\left( x_1^z-\frac{x_1^{\phi}}{R_0}\partial_\phi Z_0\right)+x_2^z, \label{eqn:Z2}
\end{equation}
\end{subequations}
where we may express their poloidal harmonic content $R_2=R_{2,0}+R_2^c\cos2\theta+R_2^s\sin2\theta$ (and similarly for $Z$) by appropriate use of multiple angle formulae. Note that the complexity has significantly increased in going to higher order, which will tend to make higher harmonic toroidal content proliferate. 
\par
Once we have these geometric transformations, we may then impose them as constraints on the Zernike basis, by simply considering the general expression in Eq.~(\ref{eqn:generalRelZerNAE}) to write,
\begin{subequations}
    \begin{gather}
        f_{2,0}^{\mathrm{NAE}}=-\left(\frac{\rho}{r}\right)^2\sum_{k=1}^\infty (-1)^kk(k+1)f_{2k,0}, \label{eqn:f20}\\
        f_{2,\pm2}^{\mathrm{NAE}}=-\left(\frac{\rho}{r}\right)^2\frac{1}{2}\sum_{k=1}^\infty (-1)^kk(k+1)f_{2k,\pm2},
    \end{gather} \label{eqn:2nd_order_constraint}
\end{subequations}
where the latter corresponds to the cosine piece for the + sign and the sine for -.

\subsubsection{Interpretation of the 2nd order transformation}
To illustrate the geometric meaning of this transformation, let us consider again the shape of the cross-section at the stellarator-symmetric point. At second order, the cross-section gains shaping beyond ellipticity \citep{landreman_figures_2021, rodriguez_magnetohydrodynamic_2023}. First they acquire some level of triangularity, $\delta$, a measure of left-right asymmetry of the cross-sections, defined as the distance between the vertical turning points of the cross-section respect to the centre along the symmetry line of the cross-section. In the near-axis description,
\begin{equation}
    \delta=2\left(\frac{Y_{2s}}{Y_{1s}}-\frac{X_{2c}}{X_{1c}}\right),
\end{equation}
where $Y$ and $X$ are along the normal and binormal, but an equivalent definition could be adapted to the shape inrespect to the lab-frame, with $Z$ instead of $Y$ and $R$ instead of $X$. 
\par
The other important ingredient is the Shafranov shift, the relative displacement of the centres of the cross-sections from one surface to the next. Following \cite{rodriguez_magnetohydrodynamic_2023}, in the normal direction (in this case the radial one as well), 
\begin{equation}
    \Delta_X=X_{20}+X_{2c}
\end{equation}
\par
The transformations in Eqs.~(\ref{eqn:R2}) and (\ref{eqn:Z2}) for the stellarator symmetric cross-section read,
\begin{subequations}
\begin{equation}
    R_2=(X_{20}+\Lambda)+(X_{2c}-\Lambda)\cos2\theta,
\end{equation}
\begin{equation}
    Z_2 = Y_{20}\Tilde{\tau}+
    \left(Y_{2s}\Tilde{\tau}-\tau^\phi Y_{1s}\left[\frac{Y_{1c}'}{2R_0}\Tilde{\tau}+\frac{X_{1c}}{2R_0}\left(\kappa_Z'-(\kappa_R+\kappa_\phi')\frac{Z_0'}{R_0}\right)\right]\right)\sin2\theta,
\end{equation}
\end{subequations}

where $\Tilde{\tau}=1/\tau^z$ and $\Lambda=Y_{1s}^2\tau^\phi[R_0(\tau^\phi-2\tau_R')+\tau^zR_0'']/4R_0^2$. From the top equation, we may read the Shafranov shift $\Delta_R$, which the transformation leaves unchanged, i.e., $\Delta_X=\Delta_R$. The triangularity, however, does change (see \cite{rodriguez2024magnetohydrodynami}) provided $\alpha\neq0$ and there is a rotation of the elliptical cross-section about the axis and across the stellarator symmetric point. An example of this is illustrated in Figure~\ref{fig:geoDefCrossSec}. There is as a result an offset in the value of triangularity when going from the axis to the lab frame. However, changing triangularity in the Frenet frame at second order directly affects triangularity in the lab-frame linearly. 

\section{Implementation of near-axis constraints}
From the previous section, we know the constraints the Fourier-Zernike coefficients representing the equilibrium in \texttt{DESC} must be subject to in order to match a prescribed near-axis behaviour. We now detail how such constraints are enforced during an equilibrium solve. To do so, we must start by being more precise with how \texttt{DESC} treats the typical fixed-boundary equilibrium problem. Using the Fourier-Zernike coefficients $\mathbf{y}=(R_{lmn}, Z_{lmn}, \lambda_{lmn})$ as degrees of freedom, it treats the problem as a constrained optimisation problem in which the cost function is the MHS force balance error, Eq.~\ref{eqn:DESCforceBalance}, 
\begin{equation}
    \mathbf{y}_\mathrm{opt} =  \underset{\mathbf{y}}{\arg\min} ~\mathbf{F}(\mathbf{y}), 
\end{equation}
subject to some imposed linear constraints ${\pmb A}\mathbf{y} = \mathbf{b}$. Let's take $N$ to be the number of coefficients, $M$ the number of truly independent constraints and take the function $\mathbf{F}$ to be evaluated on a $\{\rho,\theta,\phi\}$ grid. 
\par
Depending on the chosen constraints, the equilibrium problem solved addresses different formulations of the problem \citep{conlin_stellarator_2024}. In the traditional fixed-boundary formulation, one constrains linear combinations of $R$ and $Z$ coefficients to match the shape of a boundary at $\rho=1$. In the near-axis scenario, we constrain a different combination of coefficients to enforce a particular form of the solution near the axis, as described in the previous sections. Once the linear $M\times N$ constraint matrix $\pmb A$ is in place, a $N\times (N-M)$ auxiliary matrix $\pmb Z$ is defined such that $\pmb{A}\pmb{Z}\mathbf{v}=0~\forall~\mathbf{v}\in\mathbb{R}^{N-M}$. Any coefficient vector $\mathbf{y}$ that satisfies the constraint may then be written as $\mathbf{y}_\mathrm{opt} = \mathbf{y}_p + Z\mathbf{v}$, where $\mathbf{y}_p$ is a particular solution of $\pmb{A}\mathbf{y}_p=\mathbf{b}$. This way, optimisation may be performed unconstrained in the $N-M$ dimensional space of $\mathbf{v}$. This method is known as the \textit{feasible direction method} \citep[Sec.~12]{nocedal1999numerical}, and is used in \texttt{DESC} to impose constraints. 
\par
We thus know how to impose the constraints within \texttt{DESC}. We use the \texttt{pyQSC} \citep{pyQSC} and \texttt{pyQIC} \citep{pyQIC} codes to feed the pertinent near-axis information into Eqs.~(\ref{eqn:firstOrdTransf}) and (\ref{eqn:2nd_order_constraint}). We shall choose to impose constraints on $R$ and $Z$ (and potentially also $\lambda$) to the wanted order in $\rho$. We shall consider the implications of imposing different combinations in the following section. Alongside these near-axis constraints additional ones may also be imposed to regularise the solution, including the total enclosed toroidal flux (to limit the aspect ratio of the equilibrium), plasma pressure and net toroidal current, which we shall take to vanish for simplicity in the remainder of the text (this is in no way necessary, and the discussion below holds for arbitrary profiles). 
\par
Finally, in addition to the constraints, we also use the near-axis solution to provide an initial guess of the equilibrium to \texttt{DESC} by evaluating the near-axis field at a finite radius \citep{landreman_constructing_2019} (appropriately translated into the Fourier-Zernike basis). As formulated, \texttt{DESC} is then ran to solve for the NAE-constrained equilibrium, by minimizing force balance under the mentioned constraints. As a final sanity check, the near-axis constraints are relaxed at a final step and the equilibrium re-solved as a fixed-surface solve. We do so to check that the resulting field is truly an equilibrium.


\subsection{Verification Tests between Global Solutions and NAE}\label{sec:NAE_verification}
Before proceeding to test the above, we introduce a number of measures to assess the correct behaviour of the constrained equilibrium. Because of the final step in the construction above, we shall achieve equilibrium to the desired level of accuracy, but the relaxation of the near-axis constraints may lead to deviations from its expected asymptotic behaviour if something went wrong. That is why it is fitting to check certain aspects of the field at the final step and compare those to the near-axis expectation. Depending on the order in the expansion-in-$\rho$ considered, different quantities may be considered. 
\par
To make the comparison as impartial as possible, we do not compare field quantities that correspond to the shape of flux surfaces which we are directly enforcing by the constraint, but rather, derived ones. We define the average discrepancies between on-axis quantities as 
\begin{equation}
    \Delta f_0 = \left\langle\frac{\left|f_0^\mathrm{DESC} - f_0^\mathrm{NAE}\right|}{f^\mathrm{NAE}_0}\right \rangle_{\phi},
\end{equation}
where $\langle\dots\rangle_\phi=\int_0^{2\pi}\dots\mathrm{d}\phi/(2\pi)$. The $\texttt{DESC}$ quantities are defined as $f_0^\mathrm{DESC}=f^\mathrm{DESC}(\rho=0)$. 
\par
For assessment of the 1st order constrained fields, we consider $f = \iota,~B_0$ and $\lambda$. In a correctly constrained equilibrium we expect $\Delta\iota_0\rightarrow0$, as the rotational transform on axis is fully determined by the elliptical cross-sections about the axis \citep{landreman_direct_2018,mercier_equilibrium_1964}. However, we must be careful here, because this is only correct when $\mathbf{j}\cdot\nabla\psi=0$ exactly. Because \texttt{DESC} only achieves this approximately, as part of the optimization program, the rotational transform will have some dependence on higher order shaping $R_2$ and $Z_2$ (see Appendix~\ref{app:iota-current-formula}). Hence, $\Delta\iota_0$ shall provide both a measure of the correct implementation of the constraints, as well as the compatibility with a global solution with $\mathbf{j}\cdot\nabla\psi=0$. This potential dependence on higher orders does not apply to the magnetic field on axis, which the modulation of the ellipse areas at first order control. The third measure is a check on the behaviour of the $\theta$ coordinate in \texttt{DESC}, which by the constraints on the shaping, and in order to satisfy force-balance, should tend towards becoming a Boozer coordinate. As shown in Section \ref{sec:SFL}, this asymptotic behaviour should manifest in the stream function, and in particular on the on-axis value, Eq.~(\ref{eqn:lamba_0}). We expect these normalised quantities then to be small in a successful implementation. To compare it to something, we shall present these alongside the errors found when constructing the global equilibrium not through the near-axis contraint, but when proceeding in the standard finite radius near-axis fixed-boundary equilibrium fashion \citep{landreman_constructing_2019}. 
\par
At 2nd order, we add the comparison of $f=V''$, the so called vacuum well of the configuration \citep{greene_brief_1998,landreman_magnetic_2020}. This quantity reflects an important aspect of the MHD stability of the magnetic field, and it directly involves the 2nd order near-axis expansion. If a near-axis expansion was designed to be stable, we would like the global construction to preserve this curated feature, and thus is natural to look at $\Delta V_0''$. 
\par
In addition to these scalar measures, we shall also consider an additional quantitative diagnostic away from the magnetic axis: the scaling of the magnetic field magnitude deviations from the ideal near-axis behaviour with $\rho$. In particular, we shall look at the radial dependence of the quasisymmetry error (as measured by the normalised Boozer $f_B$ metric \citep{rodriguez_measures_2022,dudt_desc_2023})\footnote{Note that here $f_B$ is the square root of the sum of the squares of the symmetry breaking $|\mathbf{B}|$ modes normalised to the square root of the sum of squares of all $|\mathbf{B}|$ modes. It is similar in spirit to the $\hat{f}_B$ in \cite{rodriguez_measures_2022}, with a different but equally valid choice of normalization.} for quasisymmetric configurations, which we expect to scale (ideally) as $\mathcal{O}(\rho^2)$ for the first order constrained problem, and $O(\rho^3)$ when constrained to second order. Note that this scaling will only be observed if the field is sufficiently quasisymmetric at 2nd order, as one must generally allow for some deviations from exact QS at 2nd order \citep{garren_existence_1991,landreman_constructing_2019}.

\section{Near-axis constrained equilibria}\label{sec:first-order}
In this section we present a number of numerical examples of equilibria solutions constrained through different orders, which we use both as benchmark and source of discussion. 

\subsection{First order constraint}
We first consider the equilibrium solutions imposing the near-axis constraints of both 0th and 1st order; namely, Eqs.~(\ref{eqn:rAxisZern}) and (\ref{eqn:firstOrdTransf}). We present three equilibria representing different classes of optimised (omnigenous) stellarators for the benchmark: a quasi-axisymmetric (QA) and a quasi-helically symmetric (QH) configurations (near-axis constructions of the configurations presented in \cite{landreman_magnetic_2022}, which we refer to as "precise QA" and "precise QH" respectively), and a quasi-isodynamic (QI) configuration (the single field period vacuum field described in section 8.2 of \citep{plunk_direct_2019}). In all cases, we are considering stellarator symmetric, vacuum fields, though note that the constraints derived are equally valid for non-symmetric equilibria. 

\par
\begin{figure}
    \centering
    \includegraphics[width=\textwidth]{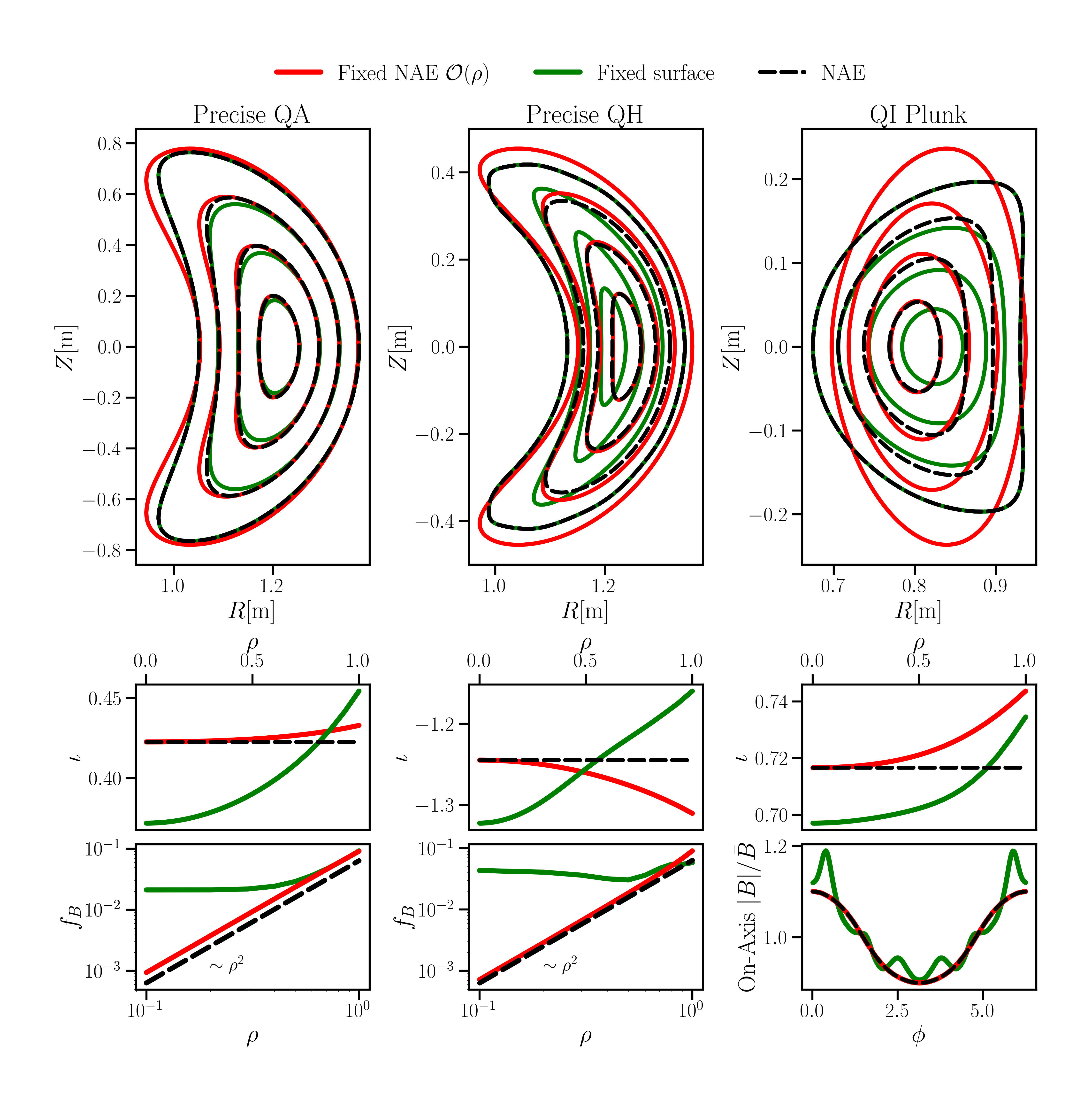}
    \caption{\textbf{Verification of first order NAE-constrained equilibria.} The figure shows a comparison of \texttt{DESC} 1st-order NAE constrained equilibrium solutions (in red) against \texttt{DESC} near-axis fixed-boundary solutions (in green) and the ideal near-axis field (broken line). From left to right the configurations correspond to the precise QA, precise QH and QI introduced in the text. (Top) Cross sections at $\phi=0$. (Middle) Rotational transform profile showing the correct matching of the NAE constrained field. (Bottom) The left plots show the Boozer quasisymmetry metric for the QS solutions (measure of $|\mathbf{B}|$ error). The expected quadratic scaling is observed in the NAE-constrained equilibria. The right plot shows the on-axis magnetic field strength for the QI solution (for which $f_B$ is not a meaingful measure).The QA and QI global equilibria were solved in \texttt{DESC} at radial ($L$), poloidal ($M$) and toroidal ($N$) spectral resolutions of $L=9,~M=9,~N=24$, and the near axis using \texttt{pyQSC} and \texttt{pyQIC}. The QH equilibrium was solved at a higher resolution of $L=10,~M=10,~N=24$, which was necessary to achieve good force balance for the fixed-surface solution.}
    \label{fig:NAE-O-rho-summary-plot}
\end{figure}
We start by summarising some of the key verification quantities in Table~\ref{tab:summary_1st_order} and Figure \ref{fig:NAE-O-rho-summary-plot}.\footnote{Each of the global equilibria in this work have average normalized force balance errors at or below 1\%, which is deemed as good numerical equilibria \citep{panici_desc_2023}.} From the comparison of cross-sections, it is clear that the NAE-constrained solutions match the NAE axis and elliptical shapes to leading order, which is not true of the fixed-boundary cases, especially for the QH and QI equilibria. These configurations are more strongly shaped than their QA counterpart, which results in a more limited radius of validity of the near-axis expansion. As a result, using the finite radius built surface as a boundary condition leads to an equilibrium that significantly departs from the original near axis behaviour. Note that the fields are being constructed at a rather low aspect ratio, which enhances this effect. This discrepancy is also apparent in the rotational transform and the behaviour of $|\mathbf{B}|$. The constrained equilibria correctly capture the on-axis rotational transform orders of magnitude better, and in the quasisymmetric fields the symmetry breaking error $f_{B}$ \citep{garren_existence_1991,dudt_desc_2023,rodriguez_measures_2022} shows the correct scaling $\sim\rho^2$. In the case of the QI configuration, the sign of a correct behaviour of $|\mathbf{B}|$ near the axis is shown by looking at its value on axis. Note that away from the axis, the global solver completes the equilibrium solution, e.g. in the case of rotational transform, introducing a small global shear. 
\par
\begin{table}
\centering
    \begin{tabular}{c|cc cc cc cc}
    \multirow{2}{*}{Configuration~} & \multicolumn{2}{c}{AR} & \multicolumn{2}{c}{$\Delta B_0$} & \multicolumn{2}{c}{$\Delta\iota_0$} & \multicolumn{2}{c}{$\Delta \lambda_0$} \\ 
     & {\tiny NAE} & {\tiny Surf} & {\tiny NAE} & {\tiny Surf}& {\tiny NAE} & {\tiny Surf}& {\tiny NAE} & {\tiny Surf} \\ \hline
 Precise QA &  2.68 &  2.68 & 2.43e-07 & 4.21e-02 & 4.95e-06 & 1.20e-01 & 3.30e-03 & 1.10 \\ \hline
 Precise QH &  4.16 &  4.31 & 2.06e-06 & 3.22e-02 & 4.95e-06 & 6.23e-02 & 2.54e-02 & 2.95 \\ \hline
 QI Plunk &  5.98 &  6.02 & 1.50e-04 & 2.53e-02 & 5.50e-05 & 2.73e-02 & 5.35e-02 & 2.58 \\ \hline
    \end{tabular}
\caption{\textbf{Quantitative verification of equilibrium field.} Table including the verification measures introduced in Sec.~\ref{sec:NAE_verification} comparing the near-axis constrained (NAE) and the fixed-boundary (Surf) equilibria. The first column shows the aspect ratio of the equilibria considered. As expected, the near-axis constrained solution performs significantly better than the fixed boundary one.}
\label{tab:summary_1st_order}
\end{table}
Although the accuracy achieved in a quantity such as $\iota_0$ is orders of magnitude better than the conventional fixed-boundary case, which would need an extremely high ($R_0/a>160$) aspect ratio to match it \citep{landreman_direct_2019}, $\Delta\iota_0$ is not down to machine precision. The imperfect force balance achieved in the equilibrium solution is likely the culprit. As indicated above, unless $\mathbf{j}\cdot\nabla\psi=0$ exactly, then the rotational transform will depend on second order shaping, which not being constrained, will affect the precision of the agreement with the NAE. The larger and more complex the higher order shaping tends to be, the easier it is for this discrepancy to be larger. This explains the relatively worse performance of the QI equilibrium, whose increased shaping can be seen in the heightened difficulty in representing the QI NAE behavior with the cylindrical toroidal angle, as explained in Appendix \ref{app:truncation}. Despite these limitations, the accuracy is practically sufficient to show that the global equilibrium behaves like the NAE near-axis. 

\subsection{Second order constraint}
Let us now repeat the construction, but this time using near-axis constructions to 2nd order. We use for that purpose the near-axis versions of the four highly quasisymmetric equilibria in \cite{landreman_magnetic_2022}. Besides precise QA and QH, which we shall now construct to 2nd order, we also consider NAE models of the other two, namely "precise QA + well” and "precise QH + well” configurations \citep{landreman_magnetic_2022}. 
We now construct NAE-constrained equilibria in \texttt{DESC}, constraining the $\mathcal{O}(\rho^2)$ $R$ and $Z$ behavior of the NAE solution. We shall also fix in this case the on-axis $\lambda$, which was found to help the equilibrium solve while not being overconstraining. We shall be particularly fixated with the \texttt{DESC} solution retaining some of the higher order features of the equilibirum. Specifically, we place renewed focus on the scaling of the quasisymmetric error and the magnetic well $V''$ \citep{landreman_magnetic_2020}. The latter is particularly important, as MHD stability is often found to be at odds with other physics and engineering objectives in stellarator optimization \citep{conlin_stellarator_2024, rodriguez_magnetohydrodynamic_2023}, but with this constraint one may enforce some measure of stability on the MHS equilibrium solution. 
\par
\begin{figure}
    \centering
    \includegraphics[width=\textwidth]{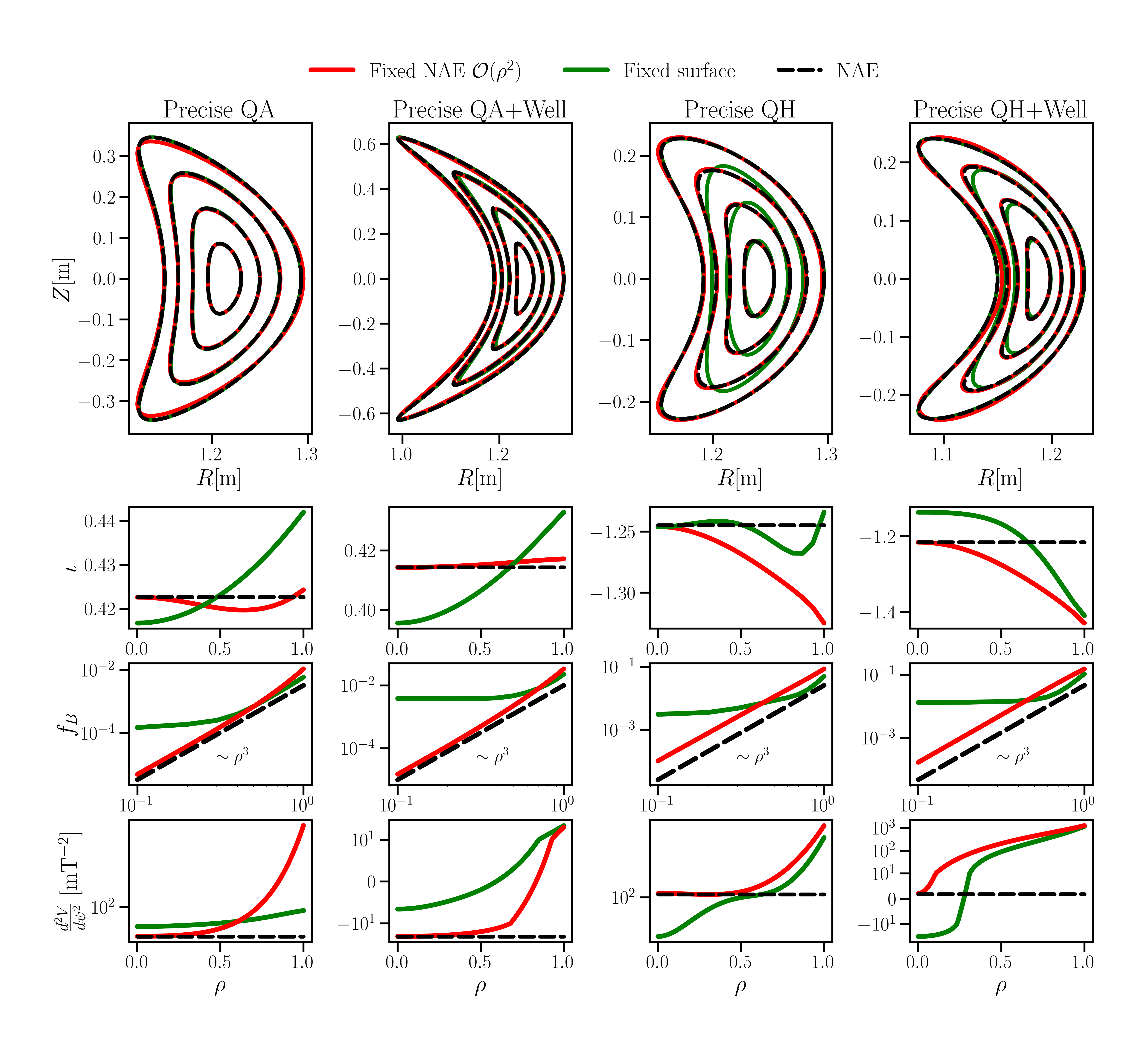}
    \caption{ 
    \textbf{Verification of second order NAE-constrained equilibria.} The figure shows a comparison of \texttt{DESC} 2nd-order NAE constrained equilibrium solutions (in red) against \texttt{DESC} fixed-boundary solutions (in green) and the ideal near-axis field (broken line). The plots correspond to the equilibria introduced in the text, the "\texttt{precise QA}" and "\texttt{precise QA+well}" on the left, the "\texttt{precise QH}" and "\texttt{precise QH+well}" on the right. (Top) Cross sections at $\phi=0$. (Middle Upper) Rotational transform profile showing the correct matching of the NAE constrained field. (Middle Lower) Boozer quasisymmetry metric for the QS solutions (measure of $|\mathbf{B}|$ error). The NAE-constrained equilibria generally adhere closely to the expected cubic scaling. (Bottom) Magnetic well parameter, where the NAE-constrained solutions are closer to the NAE value on-axis, which in the "\texttt{QA+well}" case corresponds to stability. Each global equilibrium was solved in \texttt{DESC} at radial ($L$), poloidal ($M$) and toroidal ($N$) spectral resolutions of $L=15,~M=10,~N=25$, and the near axis solutions using \texttt{pyQSC}.}
    \label{fig:O-rho-sqd-summary-plot}
\end{figure}

The results for the second order are given in Figure \ref{fig:O-rho-sqd-summary-plot} and summarized quantitatively in Table \ref{tab:summary_2nd_order}. Once again, the flux surfaces and rotational transform near the axis are better matched by the NAE-constrained equilibrium than by the fixed-surface. The Boozer quasisymmetry error ($f_{B}$) \citep{garren_existence_1991,dudt_desc_2023,rodriguez_measures_2022} scales like $\mathcal{O}(\rho^3)$, as expected for a 2nd order NAE field, highly quasisymmetric at 2nd order \citep{landreman_constructing_2019}. Finally, and most importantly, the magnetic well parameter is matched to the NAE solution to a good degree. 
\par
\begin{table}
\centering
    \begin{tabular}{c|cc cc cc cc cc}
    \multirow{2}{*}{Configuration~} & \multicolumn{2}{c}{AR} & \multicolumn{2}{c}{$\Delta B_0$} & \multicolumn{2}{c}{$\Delta\iota_0$} & \multicolumn{2}{c}{$\Delta V_0^{\prime\prime}$} \\ 
     & {\tiny NAE} & {\tiny Surf} & {\tiny NAE} & {\tiny Surf}& {\tiny NAE} & {\tiny Surf}& {\tiny NAE} & {\tiny Surf} \\ \hline
Precise QA &  6.51 &  6.51 & 1.84e-07 & 8.62e-05 & 6.28e-07 & 1.40e-02  & 3.91e-04 & 2.22e-02 \\ \hline
 Precise QA+Well &  4.84 &  4.83 & 8.41e-08 & 2.07e-03 & 6.34e-06 & 4.52e-02  & 4.00e-04 & 7.04e-01 \\ \hline
 Precise QH &  8.80 &  8.83 & 6.43e-06 & 1.89e-03 & 8.93e-04 & 1.12e-03  & 1.92e-02 & 6.23e-01 \\ \hline
 Precise QH+Well &  9.05 &  9.15 & 6.76e-07 & 8.40e-03 & 1.24e-04 & 6.50e-02  & 1.32e-01 & 1.95e+01 \\ \hline
    \end{tabular}
\caption{\textbf{Table with quantitative verification of equilibrium field.} Table including the verification measures introduced in Sec.~\ref{sec:NAE_verification} comparing the near-axis constrained (NAE) and the fixed-boundary (Surf) equilibria. The first column shows the aspect ratio of the equilibria considered. As expected, the near-axis constrained solution performs significantly better than the fixed boundary one.}
\label{tab:summary_2nd_order}
\end{table}

\section{Discussion and Conclusions}









In this work, we present a natural way to use the near-axis expansion to obtain global 3D MHS equilibria that preserve the near-axis properties. We do so by constructing geometric constraints for the Fourier-Zernike coefficients as used by the \texttt{DESC} code to solve for global equilibria. The procedure successfully reproduces the properties of near-axis fields and provides valid global equilibria, as shown through a variety of examples. In particular, it performs significantly better than the conventional fixed-boundary approach, especially at low aspect ratios. This difference is particularly successful in preserving the MHD-stability-linked magnetic well property of the fields close to the axis, a particular sore point of the typical global construction approach. 

The work presented in this paper opens the door to many applications. While the NAE-constrained global solutions obtained from these methods do not necessarily match the NAE far from axis, they do offer an excellent initial condition for conventional stellarator optimization, where at least part of the solution matches the desirable quantities from the NAE. This is a more desirable starting point in  optimization space than starting from a poor fixed-boundary equilibrium (which may result from a fixed-boundary solve using a low aspect ratio surface from the NAE, that may not even be physical). Because the NAE constraints defined in this paper do not define a unique equilibrium (i.e. the final equilibrium depends on the initial state), these constraints offer a flexibility to find solutions with similar behavior to the NAE near-axis and different desired behavior near the boundary. This could be exploited to achieve certain off-axis properties. Further investigation is warranted into the use of these near-axis constraints to aid in stellarator optimization, either as initial guesses or as constraints during optimization. 

Future work also includes the implementation of the alternative constraint formulation outlined in Appendix \ref{app:desc-theta-constraint}, which may further improve the match between the NAE and the NAE-constrained equilibria for fixed spectral resolution. Additionally, the adoption of a generalized toroidal angle in \texttt{DESC} would allow for the use of the Boozer toroidal angle directly as the computational angle, which would render the constraint simpler to implement and more exact as well, by eliminating the need for fitting of the NAE behavior to the cylindrical toroidal angle.


\section{Acknowledgements}

The authors would like to thank Matt Landreman and Rogerio Jorge for fruitful discussions. 

\section{Funding}
This work is funded through the SciDAC program by the US Department of Energy, Office of Fusion Energy Science and Office of Advanced Scientific Computing Research under contract number DE-AC02-09CH11466, DE-SC0022005, and by the Simons Foundation/SFARI (560651). The United States Government retains a non-exclusive, paid-up, irrevocable, world-wide license to publish or reproduce the published form of this manuscript, or allow others to do so, for United States Government purposes. E. R. was partially supported by a grant by Alexander-von-Humboldt-Stiftung, Bonn, Germany, through a postdoctoral research fellowship.

\section{Data Availability}
The data and script that support these findings are freely available on Princeton Data Commons \citep{panici_dario_dataset_2025} with DOI \href{https://doi.org/10.34770/tf2m-vq11}{https://doi.org/10.34770/tf2m-vq11}

\appendix

\section{Plasma current formulation of equilibrium} \label{app:iota-current-formula}
In this Appendix we briefly present the alternative plasma current formulation of the inverse-coordinate representation problem, to show that the knowledge of flux surface geometry and toroidal current profile (instead of rotational transform) is enough to uniquely specify the magnetic field. 
\par
Let us start by writing the covariant components of the magnetic field explicitly in terms of the geometry of flux surfaces. Taking projections of $\mathbf{B}$ in Eq.~(\ref{eqn:Binx}) and (\ref{eqn:covB}),
\begin{subequations}
    \begin{gather}
        B_\vartheta=\mathcal{J}_\vartheta^{-1}(g_{\phi\vartheta}+\iota g_{\vartheta\vartheta}), \label{eqn:B_vartheta}\\
        B_\phi=\mathcal{J}_\vartheta^{-1}(g_{\phi\phi}+\iota g_{\vartheta\vartheta}),
    \end{gather}
\end{subequations}
where the metric elements are defined to be $g_{q_iq_j}\stackrel{\cdot}{=}\partial_{q_i}\mathbf{x}\cdot\partial_{q_j}\mathbf{x}$, and $\{q_i\}$ represent some straight field line coordinates. 
\par
Now compute the net toroidal current $I(\psi)$,
\begin{equation}
    I(\psi)=\int_\psi \mathbf{j}\cdot\nabla \phi~\mathrm{d}^3\mathbf{r},
\end{equation}
integrating through the volume bounded by flux surface $\psi$. Writing $\mathbf{j}$ in terms of the covariant representation of $\mathbf{B}$, it is straightforward to show from the above that $\mu_0I=\langle B_\vartheta\rangle_{\vartheta\phi}$, where $\langle\dots\rangle_{\vartheta\phi}$ denotes an average over $\vartheta$ and $\phi$. Using the form for $B_\vartheta$ in Eq.~(\ref{eqn:B_vartheta}),
\begin{equation}
    \iota=\frac{\mu_0I - \langle\mathcal{J}_\vartheta^{-1}g_{\vartheta\phi}\rangle_{\vartheta\phi}}{\langle\mathcal{J}_\vartheta^{-1}g_{\vartheta\vartheta}\rangle_{\vartheta\phi}}.
\end{equation}
Thus, if one knows the geometry of flux surfaces and the current profile $I$, one may easily compute the rotational transform, and therefore the magnetic field $\mathbf{B}$. 

\par
In the case of a more general coordinate system that is not necessarily a straight-field-line one, we may write

\begin{equation}
    \iota=\frac{\mu_0 I-\langle\mathcal{J}_\theta^{-1}[(1+\partial_\theta \lambda)g_{\theta\phi}-g_{\theta\theta}\partial_{\phi}\lambda ]\rangle_{\theta\phi}}{\langle\mathcal{J}_\theta^{-1}g_{\theta\theta}\rangle_{\theta\phi}}.
    \label{eqn:iota_DESC_calc}
\end{equation}

In a vacuum field, these expressions for $I=0$ may be used to learn about the limiting behaviour of the rotational transform on axis. In the limit $\varrho\rightarrow0$, following the near-axis expansion and using the cylindrical coordinate system, the metric components $g_{\theta\theta}\sim r^2(\partial_\theta R_1^2+\partial_\theta Z_1^2)$ and $g_{\theta\phi}\sim r(R_0'\partial_\theta R_1+Z_0'\partial_\theta Z_1)$ appear to make $\iota(0)$ a function of the leading flux surface shaping. However, upon averaging over $\theta$ this leading contribution in the numerator vanishes (leaving an order $\rho^2$ term that is of the same order as the denominator). Thus, we are left with an explicit dependence of the rotational transform on axis on second order quantities. If $\mathbf{j}\cdot\nabla\psi=0$ (and not just $\langle \mathbf{j}\cdot\nabla\psi\rangle_\psi=0$, the flux surface average where $\langle \mathbf{j}\cdot\nabla\psi\rangle_\psi=\iint\mathbf{j}\cdot\nabla\psi\sqrt{g}d\theta d\phi$) which always holds true for a magnetic field with flux surfaces), then the rotational transform on axis becomes an expression that only depends on $R_1$, $Z_1$ and $\lambda_1$ once again, as one would expect from Mercier \citep{mercier_equilibrium_1964}.  

\section{Imposing the constraint on a general $\theta_\texttt{DESC}$} \label{app:desc-theta-constraint}
As we have noted in the text, both formally and in practice, imposing the constraints from the near-axis expansion directly into \texttt{DESC} using the Boozer poloidal angle has the benefit of simplicity, but forces the numerical solver to use $\theta_B$ as its poloidal coordinate. Although there is no inconsistency in doing so, in practice it can prove excessively restrictive. Representing the numerical solution in Boozer coordinates is generally not the most efficient. To leverage the freedom in choice of poloidal angle, we must then impose the near-axis constraints in an angle-agnostic way.
\par
Write,
\begin{equation}
    \theta_B=\theta_\texttt{PEST}+\iota\nu=\theta_\texttt{DESC}+\iota\nu+\lambda,
\end{equation}
where $\iota$ and $\nu$ are functions defined within the near-axis framework, and $\lambda$ is the stream function that \texttt{DESC} solves for. This mapping between the Boozer and \texttt{DESC} poloidal angles can be expanded to yield,
\begin{multline}
    \theta_B\approx\theta_\texttt{DESC}+(\iota_0\nu_0+\lambda_0)+r\left(\iota_0\nu_{1c}+\frac{\rho}{r}\lambda_{1c}\right)\cos\theta_\texttt{DESC}+\\
    +r\left(\iota_0\nu_{1s}+\frac{\rho}{r}\lambda_{1s}\right)\sin\theta_\texttt{DESC}+\dots \label{eqn:thetaB_tehetaDesc_approx}
\end{multline}
To treat the distinction between the Boozer and \texttt{DESC} poloidal angles perturbatively, we shall choose $\lambda_0=-\iota_0\nu_0$ on axis. Imposing such a constraint in \texttt{DESC} is straightforward, and in practice appears not to be excessively constraining.
\par
The description of flux surfaces $R$ and $Z$ in the near-axis expansion can with this expression for $\theta_B$ be expressed in terms of the \texttt{DESC} poloidal angle. Note that doing so implies changing the constraints introduced previously in the paper by terms that involve $\lambda$ explicitly. Because we make the poloidal angle transformation within the near-axis framework, the constraints on \texttt{DESC} variables are, to second order, still linear, thus the current framework in \texttt{DESC} may be exploited.
\par
Let us be more explicit in the construction of the constraints, and take $R$ as our reference. Write the near-axis expression,
\begin{multline}
    R=R_0+r(R_{1c}\cos\theta_B+R_{1s}\sin\theta_B)+ \\
    +r^2(R_{20}+R_{2c}\cos2\theta_B+R_{2s}\sin2\theta_B)+O(r^3).
\end{multline}
Substituting Eq.~(\ref{eqn:thetaB_tehetaDesc_approx}) into this, and collecting the leading order terms,
\begin{subequations}
\begin{align}
    R\approx & R_0+r\left(R_{1c}\cos\theta_\texttt{DESC}+R_{1s}\sin\theta_\texttt{DESC}\right) \\
    & + r^2\left[R_{20}-\frac{1}{2}R_{1c}\left(\iota_0\nu_{1s}+\frac{\rho}{r}\lambda_{1s}\right)+\frac{1}{2}R_{1s}\left(\iota_0\nu_{1c}+\frac{\rho}{r}\lambda_{1c}\right)\right] \\
    & + r^2\left[R_{2c}+\frac{1}{2}R_{1c}\left(\iota_0\nu_{1s}+\frac{\rho}{r}\lambda_{1s}\right)+\frac{1}{2}R_{1s}\left(\iota_0\nu_{1c}+\frac{\rho}{r}\lambda_{1c}\right)\right]\cos 2\theta_\texttt{DESC} \\
    & + r^2\left[R_{2s}-\frac{1}{2}R_{1c}\left(\iota_0\nu_{1c}+\frac{\rho}{r}\lambda_{1c}\right)+\frac{1}{2}R_{1s}\left(\iota_0\nu_{1s}+\frac{\rho}{r}\lambda_{1s}\right)\right]\sin 2\theta_\texttt{DESC}.
\end{align}
\end{subequations}
Thus at second order, the constraints on the poloidal modes of $R$ will simply be modified from the straightforward Boozer constraint as,
 \begin{subequations}
 \begin{align}
     \Tilde{R}_{20}=R_{20}-\frac{\iota_0}{2}\left(R_{1c}\nu_{1s}-R_{1s}\nu_{1c}\right)-\frac{\rho}{2r}\left(R_{1c}\lambda_{1s}-R_{1s}\lambda_{1c}\right), \\
     \tilde{R}_{2c}=R_{2c}+\frac{\iota_0}{2}\left(R_{1c}\nu_{1s}+R_{1s}\nu_{1c}\right)+\frac{\rho}{2r}\left(R_{1c}\lambda_{1s}+R_{1s}\lambda_{1c}\right), \\
     \tilde{R}_{2s}=R_{2s}-\frac{\iota_0}{2}\left(R_{1c}\nu_{1c}-R_{1s}\nu_{1s}\right)-\frac{\rho}{2r}\left(R_{1c}\lambda_{1c}-R_{1s}\lambda_{1s}\right).
 \end{align}
 \end{subequations}
To explicitly construct the constraints in \texttt{DESC}, we must then resolve these in $\phi$. Resolving the near axis quantities in Fourier space is numerically straightforward. However, the product with $\lambda$, which is part of the solution of \texttt{DESC} is trickier. In general, the product of two functions of $\phi$ will involve a convolution, and thus quite a large number of terms involving different modes in $\lambda$. 
\par
Let us start by being explicit about the construction of $\lambda_1$, which using the Zernike basis in \texttt{DESC} is,
 \begin{align*}
     \lambda_{1c,n}=-\sum_{k=1}^M(-1)^k k\lambda_{2k-1,1,n}, \\
     \lambda_{1s,n}=-\sum_{k=1}^M(-1)^k k\lambda_{2k-1,1,-n}. \\
 \end{align*}
 Now, consider what happens with a product of two real finite Fourier series in $\phi$. Call them $f(\phi)$ and $g(\phi)$, and let both of them include sine and cosine modes of up to $N$. Their product may be written in terms of convolutions of their coefficients as,
 \begin{multline}
     f(\phi)g(\phi)=f_{0}g_{0}+\frac{1}{2}\sum_{k=1}^N\left(f_kg_k+f_{-k}g_{-k}\right) + \\
     +\frac{1}{2}\left[\sum_{k=0}^{N-n}\left(f_{n+k}g_k+f_{-(n+k)}g_{-k}+f_{k}g_{m-k}+f_{-k}g_{-(n+k)}\right)+\right. \\
     \left.+\sum_{k=0}^n\left(f_{n-k}g_{k}-f_{-(n-k)}g_{-k}\right)\right]\cos n\phi \\
     +\frac{1}{2}\left[\sum_{k=0}^{N-n}\left(-f_{n+k}g_{-k}+f_{-(n+k)}g_{k}+f_{k}g_{-(m-k)}-f_{-k}g_{n+k}\right)+\right. \\
     \left.+\sum_{k=0}^n\left(f_{n-k}g_{-k}-f_{-(n-k)}g_{k}\right)\right]\sin n\phi,
 \end{multline}
where we should take coefficients with $|n|>N$ to vanish.

\section{Unravelling the Zernike basis} \label{app:zern_to_nae}
In this appendix we detail how one is to transform the representation of a function $f$ in the Zernike basis, $f_{ml}$ in Eq.~(\ref{eqn:fZer}), into the Taylor-Fourier basis of the near-axis expansion, $f_{nm}^{C/S}$ in Eq.~(\ref{eqn:fNAE}).

\subsection{Zeroth order connection}
Let us start with the constraint on those modes that are $\rho$-independent. In Eq.~(\ref{eqn:fNAE}), this corresponds to the coefficient $f_{00}^C$, but in Eq.~(\ref{eqn:fZer}) this is a linear combination of even modes $f_{lm}$. 
\par
Evaluating $f(\rho=0,\theta)$, and using the Jacobi polynomial
\begin{equation*}
    \mathcal{R}_{l}^0(\rho=0)=\frac{(-1)^{l/2}\frac{l}{2}!}{0!0!\frac{l}{2}!} = (-1)^{l/2},
\end{equation*}
for even $l$,
\begin{equation}
    f_{00}^C=\sum_{k=0}^\infty(-1)^kf_{2k,0}. \label{eqn:f00}
\end{equation}

\par
\subsection{First order connection}
Let us consider now $f_{11}^C$, which is a coefficient multiplying $\cos\theta$ and a single power of $\rho$. To select terms that satisfy the later, it is convenient to evaluate $\mathrm{d}\mathcal{R}_l^m/\mathrm{d}\rho$ and evaluate it at $\rho=0$, noting that only the $m=\pm 1$ modes will have the desired $\cos\theta$ or $\sin\theta$ mode. Doing so,
\begin{equation*}
    \left.\frac{\mathrm{d}\mathcal{R}_l^{|\pm1|}}{\mathrm{d}\rho}\right|_{\rho=0}=(-1)^{\frac{l-1}{2}}\frac{l+1}{2}.
\end{equation*}
Then, we simply write,
\begin{subequations}
    \begin{gather}
        f_{11}^C=-\frac{\rho}{r}\sum_{k=1}^\infty (-1)^k k f_{2k-1,1},\\
        f_{11}^S=-\frac{\rho}{r}\sum_{k=1}^\infty (-1)^k k f_{2k-1,-1}.
    \end{gather} \label{eqn:f11}
\end{subequations}
This puts on the table an important element, which is that the contribution from higher-order Zernike modes is amplified. This could become problematic as the amplification of higher orders may lead to noise-enhancement, however this has not yet been observed in practice. 
\par
\subsection{Generalised connection}
This approach can be generalised to any order easily. Taking the $j$-th derivative with $0<j<m<l$, and evaluating at $\rho=0$,
\begin{equation*}
    \left.\frac{\mathrm{d}^{j}\mathcal{R}_l^m}{\mathrm{d}\rho^{j}}\right|_{\rho=0}=(-1)^{\frac{l-j}{2}}\frac{\left(\frac{l+j}{2}\right)!j!}{\left(\frac{l-j}{2}\right)!\left(\frac{j+|m|}{2}\right)!\left(\frac{j-|m|}{2}\right)!}.
\end{equation*}
Accounting for the artificial $j!$ that arises from computing the derivatives respect to $\rho$, we may write the following expression for $f_{lm}^{C/S}$,
\begin{equation}
    f_{l|m|}^{C/S}=\left(\frac{\rho}{r}\right)^l\times\begin{cases}
    \begin{aligned}
        &(-1)^{l/2}\sum_{k=l/2}^\infty (-1)^k\binom{k+l/2}{l}\binom{l}{(l-|m|)/2}f_{2k,m} \quad |m| \mathrm{~is~even} \\
        &(-1)^{(l+1)/2}\sum_{k=(l+1)/2}^\infty (-1)^k\binom{k+(l-1)/2}{l}\binom{l}{(l-|m|)/2}f_{2k-1,m} \quad |m| \mathrm{~is~odd}, \\
        \end{aligned}
    \end{cases} \tag{\ref{eqn:generalRelZerNAE}}
\end{equation}
and the sign of $m$ (positive or negative) represents the cosine or sine (respectively) near-axis component. The above only holds for $l\geq |m|$, and $m$ sharing the same parity as $l$. Otherwise, the contribution will be zero. Note that the ratio $\rho/r$ is a constant.  
\par

\section{First Order Coefficients Derivation}\label{app:1st-order-coeff-derivation}

In this section we explicitly calculate what the first-order NAE coefficients for $R,Z$ are. A similarly explicit derivation is carried out for the second-order coefficients, however it is tedious and will not be written out in this work. We recall Eqs. \ref{eqn:R1} and \ref{eqn:Z1} here:
\begin{align}
    R_1&=x_1^R-x_1^\phi\frac{\partial_\phi R_0}{R_0}\\
    Z_1&= x_1^z-x_1^\phi\frac{\partial_\phi Z_0}{R_0},
\end{align}
where $R_0(\phi),~Z_0(\phi)$ are the $R,~Z$ of the magnetic axis that is known fully as a function of $\phi$ within the NAE. These equations yield a simple form of the first-order geometric behavior, which can be written as:
\begin{align}   
R_1&=\mathcal{R}_{1,1}(\phi)\cos\theta+\mathcal{R}_{1,-1}(\phi)\sin\theta\\
Z_1&=\mathcal{Z}_{1,1}(\phi)\cos\theta+\mathcal{Z}_{1,-1}(\phi)\sin\theta
\end{align}
The coefficients $\mathcal{R}_{1,1}, \mathcal{R}_{1,-1}, \mathcal{Z}_{1,1}, \mathcal{Z}_{1,-1}$ are calculated by recalling the form of $\mathbf{x}_1$ from the NAE:
\begin{equation}
    \mathbf{x}_1=\cos\theta(X_{1c}\hat{\kappa}+Y_{1c}\hat{\tau})+\sin\theta (X_{1s} \hat{\kappa} + Y_{1s}\hat{\tau} ).
\end{equation}
The coefficients $X_{1c}, X_{1s}, Y_{1c}, Y_{1s}$ are outputs of the NAE expansion at first-order.\footnote{Note that here these are harmonics of the poloidal angle $\theta$, and not a helical angle as is customary in numerical implementations of the near-axis such as \texttt{pyQSC}.} The Frenet-Serret basis vectors are also known, as well as their cylindrical projections, which we denote $y^j$ for $y \in \{\hat{\kappa}, \hat{\tau}, \hat{b_0}\}$ and $j \in \{R, \phi, Z\}$ denoting the cylindrical basis vector along which that Frenet-Serret basis vector is projected. With these defined, all that remains is to write the explicit forms of the projections of $\mathbf{x}_1$ and collect the terms attached to the $\cos\theta$ and $\sin\theta$ terms. Writing the projections of $\mathbf{x}_1$ explicitly:

\begin{align}
    x_1^R &= \mathbf{x}_1 \cdot \hat{R} = \cos\theta \left( X_1^c \kappa^R + Y_1^c \tau^R \right) + \sin\theta \left( Y_1^s \tau^R + X_1^s \kappa^R \right)\\
    x_1^\phi &= \mathbf{x}_1 \cdot \hat{\phi} = \cos\theta \left( X_1^c \kappa^{\phi} + Y_1^c \tau^{\phi} \right) + \sin\theta \left( Y_1^s \tau^{\phi} + X_1^s \kappa^{\phi} \right)\\ 
    x_1^z &= \mathbf{x}_1 \cdot \hat{Z} = \cos\theta \left( X_1^c \kappa^{Z} + Y_1^c \tau^{Z} \right) + \sin\theta \left( Y_1^s \tau^{Z} + X_1^s \kappa^{Z} \right)
\end{align}

Plugging the above into the expressions for $R_1$ yields:

\begin{align}
    R_1&=\mathcal{R}_{1,1}(\phi)\cos\theta+\mathcal{R}_{1,-1}(\phi)\sin\theta\\
    &=\left[X_1^c \left(\kappa^R - \kappa^{\phi} \frac{\partial_{\phi}R_0}{R_0}\right) + Y_1^c \left( \tau^R - \tau^{\phi}\frac{\partial_{\phi}R_0}{R_0}\right)   \right]\cos\theta \\
    &+ \left[Y_1^s \left(\tau^R - \tau^\phi\frac{\partial_{\phi}R_0}{R_0}\right) + X_1^s \left(\kappa^R - \kappa^\phi\frac{\partial_{\phi}R_0}{R_0}\right) \right] \sin\theta
\end{align}

Where we identify the terms multiplying the $\cos\theta$ as $\mathcal{R}_{1,1}(\phi)$ and those multiplying $\sin\theta$ as $\mathcal{R}_{1,-1}(\phi)$. The same may be carried out for $Z_1$, yielding:
\begin{align}
    Z_1&=\mathcal{Z}_{1,1}(\phi)\cos\theta+\mathcal{Z}_{1,-1}(\phi)\sin\theta\\
    &=\left[X_1^c \left(\kappa^Z - \kappa^{\phi} \frac{\partial_{\phi}Z_0}{R_0}\right) + Y_1^c \left( \tau^Z - \tau^{\phi}\frac{\partial_{\phi}Z_0}{R_0}\right)   \right]\cos\theta \\
    &+ \left[Y_1^s \left(\tau^Z - \tau^\phi\frac{\partial_{\phi}Z_0}{R_0}\right) + X_1^s \left(\kappa^Z - \kappa^\phi\frac{\partial_{\phi}Z_0}{R_0}\right) \right] \sin\theta
\end{align}

where again one may identify $\mathcal{Z}_{1,1}(\phi)$ and $\mathcal{Z}_{1,-1}(\phi)$ by matching coefficients. With these quantities known as a function of the cylindrical toroidal angle $\phi$, the final step to arrive at coefficients we may use to constrain the DESC Fourier-Zernike coefficients to enforce the first-order behavior to match the NAE is to Fourier transform these quantities in $\phi$ to finally yield $\mathcal{R}_{1,1,n}, \mathcal{R}_{1,-1,n}, \mathcal{Z}_{1,1,n}, \mathcal{Z}_{1,-1,n}$, which are then used in Eq. \ref{eqn:firstOrdTransf}.

\section{Second order flux surface description}\label{app:2nd-order-derivation}
\begin{figure}
    \centering
    \includegraphics[width=0.5\linewidth]{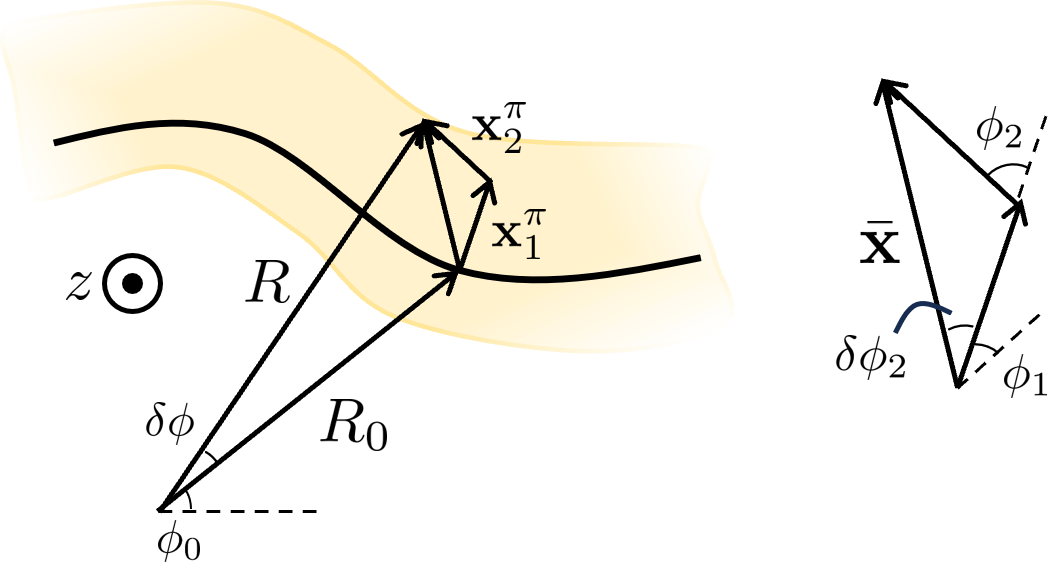}
    \caption{\textbf{Diagram illustrating the key element for the geometric transformation at second order.} Schematic diagram showing the position of a point in the surface (at radial distance $R$ and angle $\phi_0+\delta\phi$), in reference to other quantities. These include the position along the magnetic axis (radial position $R_0$ and angle $\phi_0$), and $\rho\mathbf{x}_1$ and $\rho\mathbf{x}_2$ from the axis to the point, projected onto the $R,~\phi_c$ plane (superindex $\pi$). The diagram to the right is a zoom-in of the triangle formed by $\{\mathbf{x}_1^\pi,\mathbf{x}_2^\pi,\bar{\mathbf{x}}\}$.}
    \label{fig:diagramGeoZernNAE2nd}
\end{figure}

In this Appendix we detail the derivation of the constraints related to the 2nd order near-axis behaviour. Let in that case $\mathbf{x}=\mathbf{r}_0+\rho\mathbf{x}_1+\rho^2\mathbf{x}_2$, where $\mathbf{x}_2=X_2\hat{\pmb\kappa}+Y_2\hat{\pmb\tau}+Z_2\hat{\pmb t}$, and consider an additional triangle defined by $\mathbf{x}_1$ and $\mathbf{x}_2$. We summarise the relevant geometry in the diagram in Figure~\ref{fig:diagramGeoZernNAE2nd}, an extension of Figure~\ref{fig:diagramGeoZernNAE}. 

The construction then follows identical geometric arguments to those used in the main text for the first order construction. Define an angle $\phi_2$ as the angle between $\mathbf{x}_2^\pi$ and $\mathbf{x}_1^\pi$ (the projection of the respective vectors onto the $Z=0$ plane), the angle $\delta\phi_2$ as the small angle of the triangle they form, with $\overline{x}$ the total length of the sum of $\mathbf{x}_2^\pi$ and $\mathbf{x}_1^\pi$. Using the sine and cosine rules, 
\begin{gather}
    \delta\phi_2\approx\frac{x_2^\pi}{x_1^\pi}\sin\phi_2, \\
    \overline{x}-x_1^\pi\approx x_2^\pi\cos\phi_2, \label{eqn:xbar}
\end{gather}
which are completely analogous to the expressions in Section \ref{sec:first-order-geom}. Now we must consider this $\overline{\mathbf{x}}$ piece together with $R_0$. Define $\delta\phi$ to be the angle subtended by $\overline{x}$ (see Figure~\ref{fig:diagramGeoZernNAE2nd}), and define an auxiliary angle $\overline{\delta\phi}$ so that $\overline{\delta\phi}=\delta\phi-\delta\phi_0$, where $\delta\phi_0=x_1^\phi/R_0$ is the angle correction derived at first order; that is, this barred $\delta\phi$ is the change in the angle die to the second order. From the sine rule, it follows
\begin{equation}
    \overline{\delta\phi} \approx \frac{x_2^\phi}{R_0}-\frac{x_1^Rx_1^\phi}{R_0^2}. \label{eqn:overline_deltaphi}
\end{equation}
For the position $R$, we also define a second order correction $\overline{\delta R}=R-R_0=\delta R_0$, where $\delta R_0=x_1^R$ is the first order form. Using $\Bar{x}$ in Eq.~(\ref{eqn:xbar}), as well as $\overline{\delta\phi}$ in Eq.~(\ref{eqn:overline_deltaphi}), and applying the cosine rule
\begin{equation}
    \overline{\delta R}\approx x_2^R+\frac{(x_1^\phi)^2}{2R_0}.
\end{equation}
Recall that this is not the end of the story, as we must put everything as a function of the cylindrical angle $\phi_c$ evaluated at the flux surface point. But all quantities we know are functions of the position along the magnetic axis. Keeping the next order correction,\footnote{One must not forget to expand the cylindrical angle correction as well }
\begin{align}
    R(\phi_c)&\approx R_0(\phi_c-\delta\phi_0+\delta\phi_0\partial_\phi\delta\phi_0-\overline{\delta\phi})+\delta R_0(\phi_c-\delta\phi_0) +\overline{\delta R}(\phi_c)\nonumber\\
    &\approx\dots+\rho^2\left[\frac{(\delta\phi_0)^2}{2}\partial_\phi^2 R_0-\overline{\delta\phi}\partial_\phi R_0-\delta\phi_0\partial_\phi \delta R_0+\delta\phi_0\partial_\phi R_0\partial_\phi\delta\phi_0+\overline{\delta R}\right].
\end{align}
Thus,
\begin{multline}
    R_2=-\frac{1}{2}\partial_\phi^2R_0\left(\frac{x_1^\phi}{R_0}\right)^2-\left(\frac{x_2^\phi}{R_0}-\frac{x_1^Rx_1^\phi}{R_0^2}\right)\partial_\phi R_0\\
    -\frac{x_1^\phi}{R_0}\partial_\phi\left(x_1^R-\frac{x_1^\phi}{R_0}\partial_\phi R_0\right)
    +\left(x_2^R+\frac{(x_1^\phi)^2}{2R_0}\right), \label{app:eqn:R2}
\end{multline}
which has a simple $\theta$-harmonic form $R_2=R_{2,0}+R_2^c\cos2\theta+R_2^s\sin2\theta$. 
\par
In the case of $Z$,
\begin{equation}
    Z_2=-\frac{1}{2}\partial_\phi^2 Z_0\left(\frac{x_1^\phi}{R_0}\right)^2-\left(\frac{x_2^\phi}{R_0}-\frac{x_1^Rx_1^\phi}{R_0^2}\right)\partial_\phi Z_0-\frac{x_1^\phi}{R_0}\partial_\phi\left( x_1^z-\frac{x_1^{\phi}}{R_0}\partial_\phi Z_0\right)+x_2^z. \label{app:eqn:Z2}
\end{equation}
Once again, we may write the $Z$ content as $Z_2=Z_{2,0}+Z_2^c\cos2\theta+Z_2^s\sin2\theta$, and find the coefficients using some algebraic manipulator.

\section{Finite-n Truncation of NAE Constraint}\label{app:truncation}
\begin{figure}
\centering
\includegraphics[width = \textwidth]{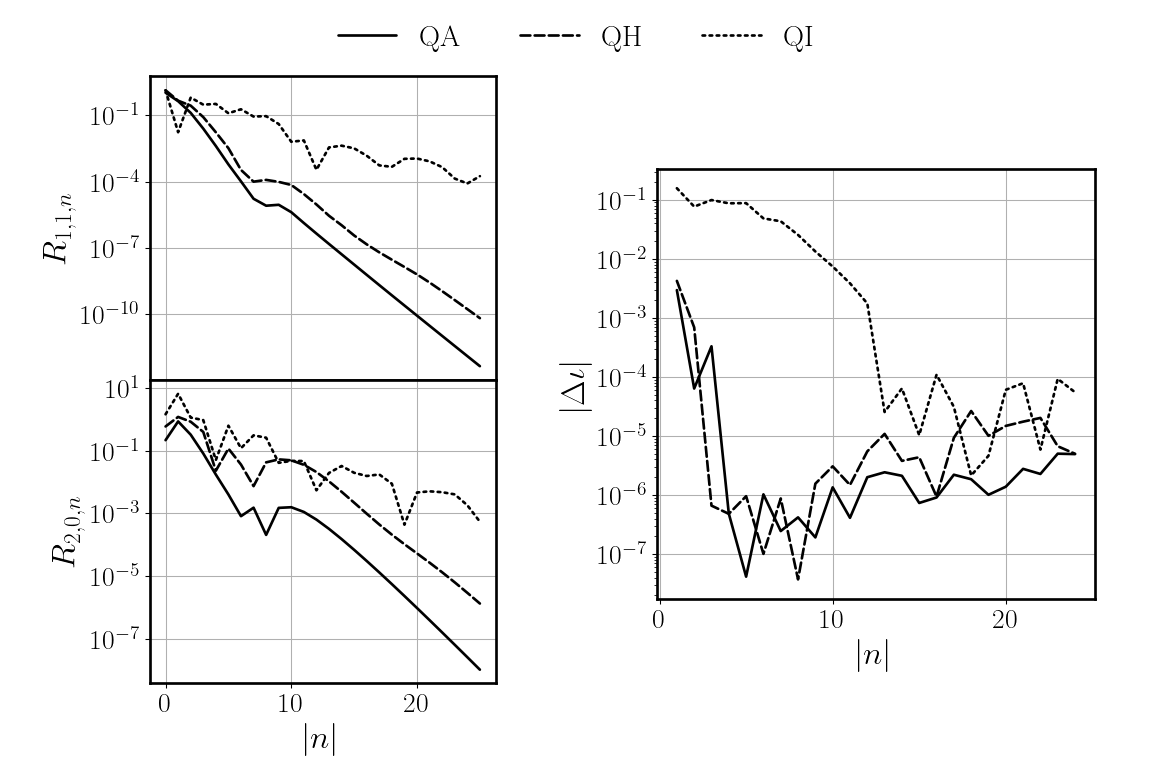}
\caption{(Left) Fourier coefficients as a function of toroidal mode number of the $O(\rho)$ NAE coefficient $R_{1,1}$ (top) and order $O(\rho)$ coefficient $R_{2,0}$ (bottom) from Eq.~\eqref{eqn:firstOrdTransf} and Eq.~\eqref{eqn:f20} respectively. (Right) Error in the NAE-constrained \texttt{DESC} solution's rotational transform on axis as a function of the toroidal resolution of the constraint used.}
\label{fig:coeff_conv}
\end{figure}

In this Appendix we present some details regarding the role of the finite toroidal resolution used to represent the equilibrium in \texttt{DESC}. Given the finite resolution of the \texttt{DESC} solver, the solution is only represented by a finite number of toroidal and poloidal modes, $N$ and $M$ respectively. When it comes to imposing the near-axis constraints as discussed in this paper, the finite resolution in the toroidal angle is particularly intrusive, as it leads to having to discard part of the Fourier content of the near-axis field. If the high-$n$ content in the shaping of the near-axis solution is significant, one then expects the constraint to deviate from the full near-axis behaviour. Such high-$n$ behaviour could be driven by a significantly shaped axis geometry, but also by the non-linear nature of the near-axis system of equations. We present an example of the magnitude of the toroidal content of some near-axis shaping coefficients to the left of Figure~\ref{fig:coeff_conv}.
\par
Even though any smooth periodic function must have a decaying toroidal spectrum that decays exponentially for sufficiently large $n$ \citep[Corollary~2.4]{stein2011fourier}, they do so at markedly different rates at finite mode number. The slower the decay in $n$, the harder it is to represent the near-axis solution in the pseudo-spectral choice of \texttt{DESC}. From the figure it can be appreciated that the decay is particularly slow for the QI field. The hardship in representing QI solutions in cylindrical coordinates is well recognised in the literature, as these configurations feature axes with straight sections that are difficult to represent with $\phi$ \citep{plunk_direct_2019, rodriguez_near-axis_2024}). In addition, it is also clear that convergence of second-order behavior is markedly slower. This is to be expected, as additional $\phi$-derivatives occur every time one moves an order up in the near-axis construction. Allowing an arbitrary computational toroidal angle $\zeta$ could help in condensing the Fourier spectra of these behaviors, as well as adopting alternative representations \citep{hindenlang2024generalized, hindenlang_computing_2025}, however this solution will not be pursued in this work. 
\par
To assess the importance of this truncation we perform the following numerical experiment: for a toroidal resolution of $N$, take the lower $n\leq N$ $\mathcal{O}(\rho)$ constraints and perform an equilibrium solve\footnote{That is, perform a minimization of force error while using the first-order NAE constraints on $R$ and $Z$, then perform a final fixed-boundary solve of the resulting solution.}. The resulting $\Delta\iota_0(n)$ is shown on the right of Figure \ref{fig:coeff_conv}. The quasisymmetric cases converge quite rapidly ($n\sim5$), suggesting in these scenarios this to be a relatively minor problem (at least at first order). In the QI case, the requirement is clearly stronger, ratifying the complications arising from the poor representation of the field. Note however that the saturation of the solution must then come from the approximate form of force-balance. In fact, the error does not even behave in a monotonic fashion, although it remains below a more than satisfactory level.


\bibliographystyle{jpp}
\bibliography{main.bib}

\end{document}